\documentclass[11pt]{article} 

\usepackage[utf8]{inputenc} 

\usepackage{geometry} 
\usepackage{graphicx} 

\usepackage{booktabs} 
\usepackage{array} 
\usepackage{paralist} 
\usepackage{verbatim} 
\usepackage{subfig} 
\usepackage{graphicx,lscape}
\usepackage{latexsym}
\usepackage{epsfig}

\usepackage{fancyhdr} 
\usepackage{sectsty}
\allsectionsfont{\sffamily\mdseries\upshape} 

\usepackage[nottoc,notlof,notlot]{tocbibind} 
\usepackage[titles,subfigure]{tocloft} 

\title{GRB 090313 and the Origin of Optical Peaks in GRB Light Curves: Implications for Lorentz Factors and Radio Flares}

\author{A. Melandri$^{1}$\thanks{E-mail: axm@astro.livjm.ac.uk}, S. Kobayashi$^{1}$, C.G. Mundell$^{1}$, C. Guidorzi$^{2,3,1}$, A. de Ugarte Postigo$^{4,3}$, \and G. Pooley$^{5}$, M. Yoshida$^{6}$, D. Bersier$^{1}$, A.J. Castro-Tirado$^{7}$, M. Jel\'{i}nek$^{7}$, A. Gomboc$^{8,10}$, \and J. Gorosabel$^{7}$, P. Kub\'{a}nek$^{7,9}$, M. Bremer$^{11}$, J.M. Winters$^{11}$, I.A. Steele$^{1}$, R.J. Smith$^{1}$, \and I. de Gregorio-Monsalvo$^{4,12}$, D. Garc\'{i}a-Appadoo$^{4}$, A. Sota$^{7}$, A. Lundgren$^{4}$ \and \\
$^{1}$ Astrophysics Research Institute, Liverpool J.Moores University, Birkenhead, CH41 1LD, UK \and
$^{2}$ Dipartimento di Fisica, Universit\`a di Ferrara, via Saragat 1, I-44100 Ferrara, Italy \and 
$^{3}$ INAF Osservatorio Astronomico di Brera, via Bianchi 46, 23807 Merate (LC), Italy \and 
$^{4}$ European Southern Obs., Casilla 19001, Santiago 19, Chile \and 
$^{5}$ Mullard Radio Astronomy Obs., Cavendish Lab., Madingley Road, Cambridge CB3 0HE, UK \and
$^{6}$ Okoyama Astrophysical Obs., National Astronomical Obs., Okayama 719-0232, Japan \and
$^{7}$ Instituto de Astrofis\'{i}ca de Andaluc\'{i}a, Glorieta de la Astronom\'{i}a s/n, E-18080 Granada, Spain \and
$^{8}$ Faculty of Mathematics and Physics, University of Ljubljana, SI-1000 Ljubljana, Slovenia \and
$^{9}$ Image Processing Lab., University of Valencia, Poligono la Coma s/n, E-46980 Valencia, Spain \and
$^{10}$ Centre of Excellence SPACE-SI, Askeroeva cesta 12, SI-1000 Ljubljana, Slovenia \and
$^{11}$ Institut de Radioastronomie Millimetrique, 38406 St. Martin d'Heres, France \and
$^{12}$ ALMA, Avda. Apoquindo 3846, Piso 19, Edificio Alsacia, Las Condes, Santiago, Chile}

\date{} 

\begin{document}
\maketitle

\begin{abstract} 

We use a sample of 19 Gamma Ray Bursts (GRBs) that exhibit single-peaked optical light 
curves to test the standard fireball model by investigating the relationship between the time 
of the onset of the afterglow and the temporal rising index. Our sample includes GRBs and 
X-ray flashes for which we derive a wide range of initial Lorentz factors ($40 < \Gamma < 450$). 
Using plausible model parameters the typical frequency of the forward shock is expected 
to lie close to the optical band; within this low typical frequency framework, we use the 
optical data to constrain $\epsilon_e$ and show that values derived from the 
early time light curve properties are consistent with published typical values derived from other 
afterglow studies. We produce expected radio light curves by predicting the temporal 
evolution of the expected radio emission from forward and reverse shock components, 
including synchrotron self-absorption effects at early time. Although a number of the GRBs 
in this sample do not have published radio measurements, we demonstrate the effectiveness 
of this method in the case of {\it Swift} GRB~090313, for which millimetric and centrimetric observations 
were available, and conclude that future detections of reverse-shock radio flares with new radio 
facilities such as the EVLA and ALMA will test the low frequency model and provide 
constraints on magnetic models.

\end{abstract} 
 
 
\section{Introduction} 
 
 With the advent of rapid optical follow-up observations of Gamma Ray
 Bursts (GRBs) (e.g. Mundell et al. 2010, Rykoff et al. 2009), the confirmed 
 lack of bright optical flashes from  most GRBs challenges a key prediction 
 of the standard fireball model in  which a reverse shock should produce
 bright, short-lived optical emission at early time (M\'esz\'aros \&
 Rees 1999; Sari \& Piran 1999; Kobayashi 2000). 
Although the lack of optical flash could be partially due to  
late observations which are not prompt enough to catch early flashes, 
it is not trivial how to explain events like GRB 090313 which exhibits
 the onset of afterglow without signatures of optical flash.  

At early time, reverse shock emission should dominate optical band
and a bright optical peak is expected to be observed when a fireball
starts to be decelerated. However, a distinctive reverse shock component 
is detected only in a small fraction of GRBs (Melandri et al. 2008). Several 
afterglows show a fattening in the light curves, interpreted as the signature 
of the rapid fading of reverse shock combined with the gradual dominance 
of forward shock emission (Akerlof et al. 1999; Sari \& Piran 1999). 
Afterglow modeling of such flattening cases implies that the magnetic
energy density in a fireball, expressed as a fraction of the equipartition 
value  of shock energy, is much larger than in the forward shock (but it still 
suggests a baryonic jet rather than a Poynting-flux dominated  jet: Fan et al. 2002; 
Zhang et al. 2003; Kumar \& Panaitescu 2003; Gomboc et al. 2008). Polarization 
measurements in a rapid decay phase of GRB 090102 afterglow shows the 
existence of large-scale magnetic fields in the revere shock region (Steele 
et al. 2009\footnote{Mundell et al. 2007b found no ordered magnetic fields 
or a very high magnetic energy density in the ejecta of GRB 060418. More 
observations are needed to give a strong conclusion on the nature of the 
ejecta (baryonic versus Poynting flux dominated) and the distribution of 
magnetization degree.}). The lack of optical flashes in most GRBs may be 
due to extreme magnetic field properties, either high magnetic energy densities 
that suppress the reverse shock  (Gomboc et al. 2008; Mimica et al. 2009) or 
very low magnetic energy densities that cause  shock  energy to be radiated 
at higher frequencies than the optical  band due  to synchrotron self-Compton
processes (Beloborodov 2005; Kobayashi et al 2007; Zou et al 2009). 
Alternatively the light curve flattening could be the result of refreshed shocks 
and episodes of energy injection (Rees $\&$ M\'esz\'aros 1998, Melandri et al. 2009).

A more conventional model would imply that the reverse shock emits photons at
frequencies much lower than the optical band. Synchrotron emission is
known to be sensitive to the properties of emitter. 
Within this framework, which we term the {\em
 low-frequency model}, a single peak in the early time optical light
 curve is produced when both of the typical synchrotron frequencies of
 forward  and reverse shock lie below the optical band (Mundell et al. 2007a); 
 the single  peak actually consists of photons equally contributed from 
 forward  and reverse  shock, the peak time represents the 
 the deceleration of a fireball and hence it provides a direct estimate
 of the initial Lorentz factor. 

In this paper, we discuss the lack of optical flashes in the context of
the low-frequency model. GRB 090313 is a typical case of a burst that
displays a rising and falling light curve, little temporal
structure, no strong spectral evolution and well-monitored
multi-wavelength behavior from early times. Here, we analyse its
multi-wavelength properties, place it into the wider context of GRBs
with single optically peaked light curves and use the characteristics of
the full sample to test the low-frequency model and its predations for 
radio light curve evolution. 
Throughout the paper we use the following conventions: the power-law
flux density is given as $F(\nu,t)\propto t^{-\alpha} \nu^{-\beta}$,
where $\alpha$ is the temporal decay index and $\beta$ is the spectral
slope; a positive value of $\alpha$ corresponds then to a decrease in
flux, while a negative value indicates an increasing in time of the
observed flux. We assume a standard cosmology with $H_0 =
70$~km~s$^{-1}$~Mpc$^{-1}$, $\Omega_{m} = 0.3$, and $\Omega_{\Lambda}=
0.7$; and all uncertainties are quoted at the $1\sigma$ confidence level
(cl), unless stated otherwise.

\section{Observations} 
 
On 2009 March 13 at 09:06:27 UT (=T0) the Burst Alert Telescope (BAT;
Barthelmy et al. 2005)  onboard {\it Swift} triggered on GRB 090313 (Mao et al. 2009a). The
BAT light curves showed a series of multiple peaks with the emission starting before 
T0-100 s and a T$_{90}$ in the 15-350 keV band starting at $\sim$ T0-3.9 s for a 
total duration of $78 \pm 19$ s (Mao et al. 2009b).  

Spectroscopic observations performed with the Gemini South telescope
provided a redshift of z=3.375 for GRB 090313 (Chornock et al. 2009b), later 
confirmed by from VLT with FORS (Th\"{o}ene et al. 2009) and X-shooter (de 
Ugarte Postigo et al. 2010; who derive a refined redshift value of 3.3736 +/- 
0.0004) observations. The estimated redshift for this afterglow confirmed again 
that the near object reported by Berger (2009) is indeed too bright to be the host 
galaxy of GRB 090313. Most likely this extended object is one of the two absorbing
systems spectroscopically detected (at redshift z=1.96 or z=1.80) along the line 
of sight of GRB 090313 (de Ugarte Postigo et al. 2010).  Radio observations 
performed with the AMI Large Array (Pooley 2009abc), the VLA (Frail $\&$ 
Chandra 2009) and the WSRT (van der Horst $\&$ Kamble 2009ab) confirmed 
the detection and fading nature of the afterglow.  

This event displayed an average $\gamma$-ray fluence of $\sim$ 1.4
$\times$ 10$^{-6}$ erg cm$^{-2}$ (Mao et al. 2009b). The redshift of the
burst (correspondent to a luminosity distance of $\sim 2.9 \times
10^{4}$ Mpc) resulted in an isotropic energy estimate of $\sim 3.4
\times 10^{52}$~ergs in the 15--150 keV observed bandpass.  
  
\subsection{{\it Swift}/XRT and {\it Swift}/UVOT data} 

Due to Moon distance observing constraints there were no prompt XRT (Burrows et al. 2005) and UVOT (Roming et al. 2005) observations.  Follow-up observations of the BAT error circle were possible only after $\sim$ 27 ks, showing a power-law decay in the X-ray (Mao $\&$ Margutti 2009) and a possible marginal detection in the UVOT-v and UVOT-b filters (Schady et al. 2009, Mao et al. 2009b). 

\subsection{Optical and Infrared data} 
 
The optical afterglow was discovered by the KAIT telescope (Chornock et
al. 2009a) and later confirmed by the GROND telescope at equatorial
coordinates (J2000) R.A. = 13$^{h}$13$^{m}$36.21$^{s}$; Dec =
+08$^{\circ}$05$^{'}$49.2$^{''}$ (Updike et al. 2009). 
The 2-m Faulkes Telescope North (FTN)  observed the optical afterglow of
GRB 090313 starting from 168 s after the burst (corresponding to 38 s in
the rest frame). Observations continued up to several weeks after the
burst with FTN, the 2-m Liverpool Telescope (LT) and the 2-m Faulkes
Telescope South (FTS) (see Table \ref{obslog0}). Late time observations
were also performed in order to better correct the entire data set from the 
contribution of the nearby object, close to the position of the afterglow. 
This object was found to have a constant flux equal to $\sim 1 \%$ of the 
peak flux of the optical afterglow, not affecting the shape of the light curve 
at early time.

\begin{figure*}
   \centering
   \includegraphics[height=9.5cm,width=10.5cm]{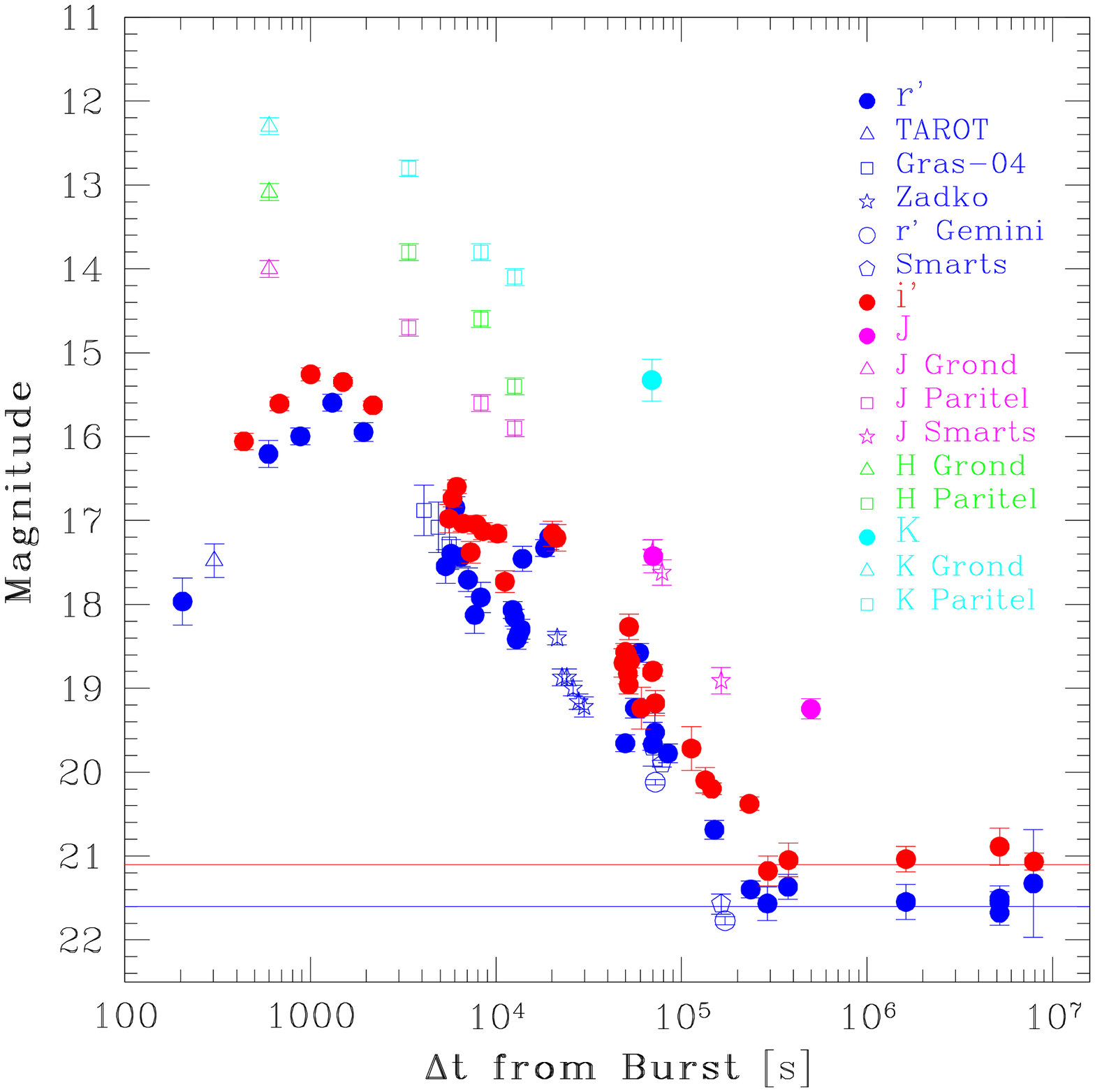} 
   \caption{Multi-band light curve in the observed frame for GRB 090313. Filled symbols are out data while GCN data (open symbols) are from: Klotz et al. 2009 (TAROT), Nissinen 2009 (Gras-04), Vaalsta $\&$ Coward 2009 (Zadko), Perley 2009 and Perley at al. 2009 (Gemini), Cobb 2009 (SMARTS), Updike et al. 2009 (GROND), Morgan et al. 2009 (PAIRITEL).}
   \label{figLC}
\end{figure*}

The optical afterglow was observed also with the 1.5m telescope at the
Observatorio de Sierra Nevada (OSN), the 0.8m IAC telescope, the 1.23m
telescope at the Calar Alto Astronomical Observatory (CAHA) and the 0.5m
Mitsume telescope in the optical bands (R and I), plus the 2.5m Nordic
Optical Telescope (NOT) and the 3.5m CAHA telescope in the near infrared
bands (J and K). It was then possible to build the light curve for all
the filters as shown in Fig.\ref{figLC}. A log of the observations is
given in Table \ref{obslog0}, where we report the mid time, integration
time, magnitude and fluxes for all our detections at different
wavelengths. Afterglow detections reported in GCNs are also shown in
Fig.\ref{figLC}. 

The optical data were calibrated using a common set of selected
catalogued stars present in the field of view. SDSS catalogued stars
were used for $r'$ and $i'$ filters, while USNO-B1 $R2$ and $I$
magnitudes have been used for the $R$ and $I$ filters respectively. $J$
and $K$ observations were calibrated with respect to the 2MASS
catalog. Next, the calibrated magnitudes were corrected for the Galactic
absorption along the line of sight ($E_{B-V} = 0.028$ mag; Schlegel et
al. 1998); the estimated extinctions in the different filters are $A_R$
$\sim$ $A_{r'}$ = 0.074 mag, $A_I$ $\sim$ $A_{i'}$ = 0.054 mag, $A_J$ =
0.025 mag, $A_H$ = 0.016 mag and $A_K$ = 0.010 mag. Corrected magnitudes
were then converted into flux densities, $F_{\nu}$ (mJy), following
Fukugita et al. (1996). Results are summarized in Table \ref{obslog0}.  
 
\subsection{Radio, mm and sub-mm data}

Continuum observations at 870 $\mu$m were carried out using LABOCA bolometer 
array, installed on the Atacama Pathfinder EXperiment (APEX$\footnote{This work is 
partially based on observations with the APEX telescope. APEX is a collaboration 
between the Max-Plank-Institut f\"ur Radioastronomie, the European Southern Observatory
and the Onsala Space Observatory.}$) telescope. Data were 
acquired on 2009 March 17 and 24 during the ESO program 082.F-9850A, under 
good weather conditions (zenith opacity values ranged from 0.24 to 0.33 at 870$\mu$m). 
Observations were performed using a spiral raster mapping,  
providing a fully sampled and homogeneously covered map in an area of diameter 
$\simeq$12$'$,  centered at the coordinates of the optical afterglow of GRB 090313. 
The total on source integration time of the two combined epochs was $\simeq$ 4.6 hours. 
Calibration was performed using observations of Saturn as well as CW-Leo, B13134, 
G10.62, and G5.89 as secondary calibrators. The absolute
flux calibration uncertainty is estimated to be $\simeq$ 11\%. The telescope pointing
was checked every hour, finding an rms pointing accuracy of 1.8$^{\prime\prime}$. Data 
were reduced using the BoA and MiniCRUSH softwares. Finally, the individual maps 
were co-added and smoothed to a final angular resolution of ~27.6$^{\prime\prime}$. We 
obtained a 3$\sigma$ detection upper limit of 14 mJy for each of the two epochs.

The radio afterglow of GRB 090313 was successfully detected by the AMI
Large Array $\sim$ 2.8 days after the burst (Pooley 2009a) and then
monitored up to $\sim$ 47 days (Pooley 2009bc) as reported in Table
\ref{obslog1}. After an initial upper limit at $\sim$ 1.7 days (van der
Horst $\&$ Kamble 2009a) a detection was reported also by the
Westerbork Synthesis Radio Telescope at $\sim$ 7.6 days (WSRT, van der
Horst $\&$ Kamble 2009b) and by the Very Large Array at $\sim$ 5.9 days
(VLT, Frail $\&$ Chandra 2009). In the mm band the afterglow was
detected with CARMA about one day (Bock et al. 2009) and then monitored
with  the Plateau de Bure Interferometer (PdBI) up to $\sim$ 20 days
after the burst event. The radio observations are reported in Table
\ref{obslog1} where the original frequency range of the observation has
been specified. 

\section{Results} 
 
 \subsection{BAT spectral and temporal analysis}
 
\begin{figure*}
    \includegraphics[height=8.5cm,width=10.5cm]{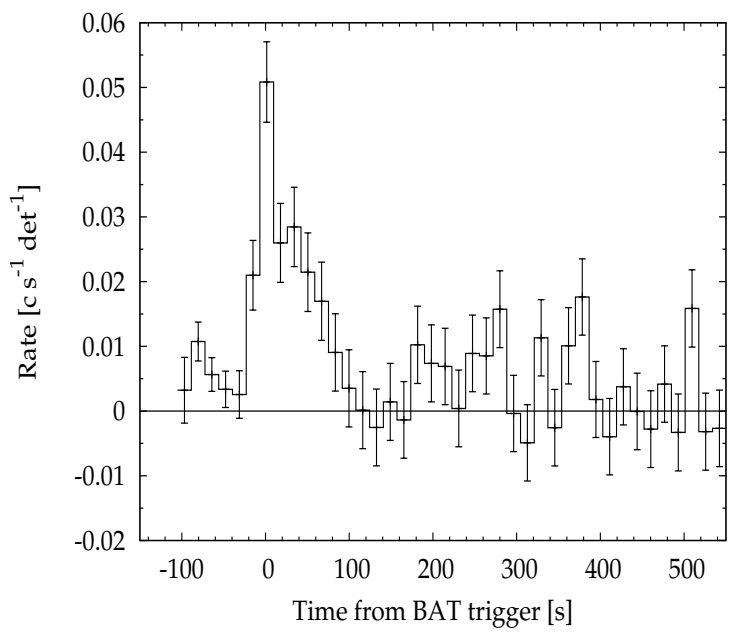} 
    \caption{Gamma-rays 16 s binning light curve for GRB 090313 as observed by BAT.}
    \label{batlc}
 \end{figure*}

We re-binned the BAT light curve of GRB 090313 with dt bins of 16.384 s in order
to better appreciate the long faint tail visible up to 500 s after the burst onset. As
reported also by Mao et al. (2009b), the mask-weighted light curve (shown
in Fig. \ref{batlc}) displays a series of multiple peaks extending long
after t=T$_{90}$ at a much fainter level. The time-averaged spectrum is
best fitted by a simple power-law model with a photon index of 1.91
$\pm$ 0.29 (Mao et al. 2009b). 

\subsection{Optical/X-ray light curve} 

Observations performed with the Faulkes North Telescope, beginning $\sim$
170 s after the burst, showed the optical afterglow rising to a maximum at
$\sim$ 1 ks (Guidorzi et al. 2009). The peak was followed by a decay with windings
and flares (possibly due to the interaction with the circum-burst
material or late time central engine activities). Around $3\times 10^5$ s,
the magnitude became constant in each filter, revealing the
presence of an underlying object at the position of the optical
afterglow. This faint ($r'$ = $21.6 \pm 0.2$ and $i'$ = $21.1 \pm 0.2$) and
apparently extended object is only 2.3" away from the optical afterglow
as reported by Berger (2009). It was not possible to separate the
contributions from the two objects in the late-time co-added
observations.  

\begin{figure*}
    \includegraphics[height=10.5cm,width=15.5cm]{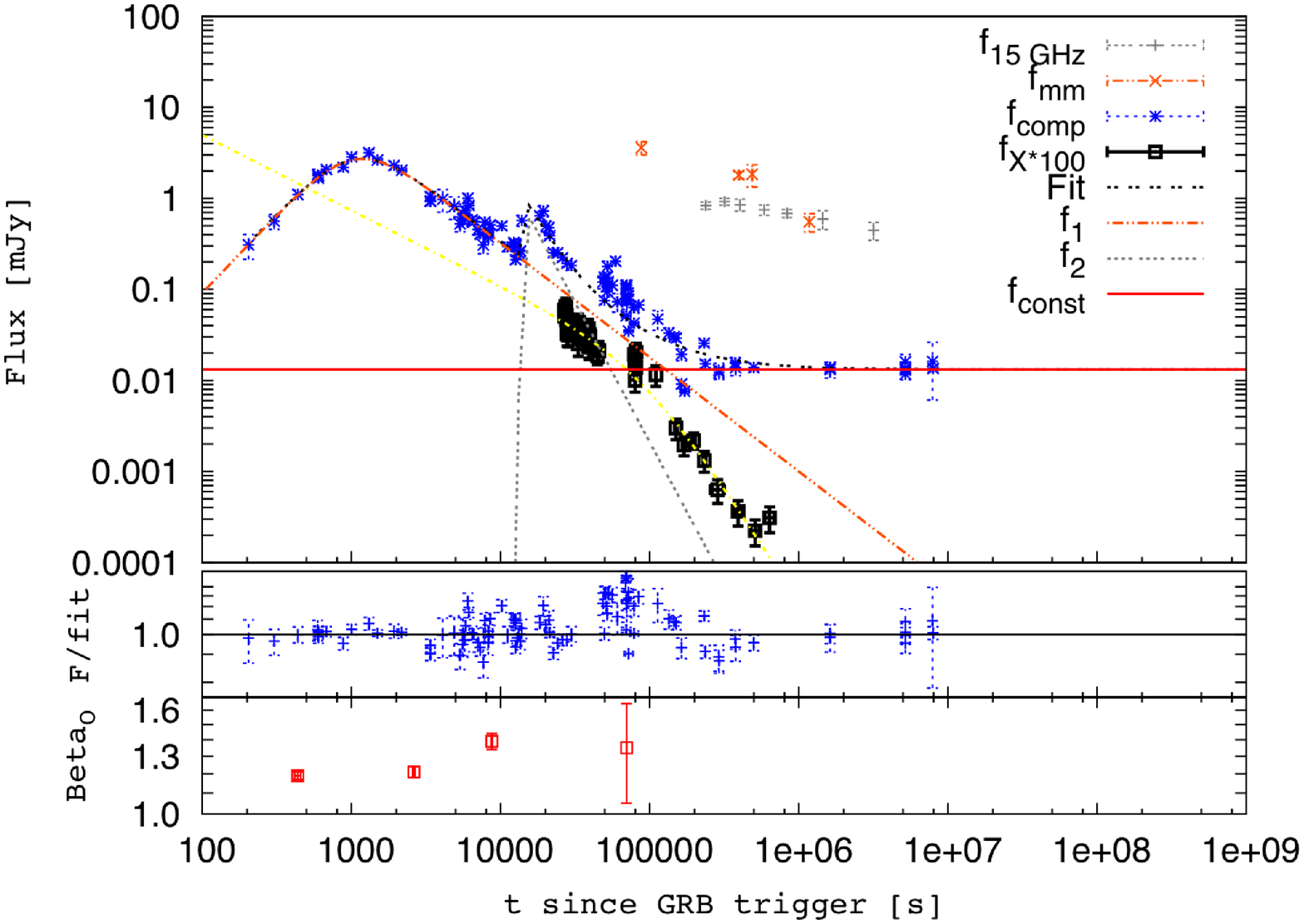} 
    \caption{Composite optical/infrared light curve (blue symbols) for the afterglow of GRB 090313. The X-ray light curve (black symbols) is well fitted by a broken power-law. See the text for the details about the fit of the composite optical/infrared light curve. In the lower panel the evolution of the optical spectral index is shown.}
    \label{figLC2}
 \end{figure*}

We model the optical light curve with a broken power-law (to fit the
peak up to $\sim 10^{4}$ s) plus an additional component to model the bumps
visible after $\sim 1.4 \times 10^{4}$ s and a constant flux to model
the behavior at late times. The fit to the component representing the
optical peak at early time gives: $\alpha_{\rm rise} = -1.72 \pm 0.41$,
$\alpha_{\rm decay} = 1.25 \pm 0.08$ and t$_{\rm peak} = 1060.9 \pm
153.6$ s. For completeness the parameters of the component modeling the
sharp bump around $\sim 10^{4}$ s are:  $\alpha_{\rm r,bump} = -83.8 \pm
8.4$, $\alpha_{\rm d,bump} = 3.0 \pm 0.8$ and t$_{\rm peak,bump} =
(14.0 \pm 0.3) \times 10^{3}$ s, (t/dt)$_{\rm peak} \sim 1$
($\chi^{2}$/dof = 769.4/77 $\sim$ 9.9). The high $\chi^{2}_{\rm red}$
for the optical fit is clearly driven by the uncertainty of the bump fit
and the variability of the data around $\sim 10^{5}$ s. However this
does not affect the goodness of the fit for the smooth early time
behavior, where the peak (rise and fall) is well constrained with
negligible variability as shown in Fig. \ref{figLC2}. 

Our independent analysis shows that the X-ray light curve of GRB 090313
is well fitted by a simple broken power-law with $\alpha_{\rm 1} = 0.83
\pm 0.49$, $\alpha_{\rm 2} = 2.56 \pm 0.46$ and $t_{\rm break} \sim 9
\times 10^{4}$ s ($\chi^{2}$/dof = 43.17/43 $\sim 1.0$). The estimated values 
for $\alpha_{\rm 1}$ and $\alpha_{\rm 2}$ could be the result of flares activity,
and the subsequent cessation, in the early XRT data. The X-ray light
curve and its fit are shown in Fig. \ref{figLC2} together with the
composite optical/infrared light curve. As we will explain in Section 3.4, the latter
has been built by re-scaling all the filters with respect to the
SDSS $i'$ band. On the bottom panel of this figure we show the
no-evolution of the optical spectral index $\beta_{\rm O}$ as derived from the
fit of the spectral energy distribution.

 \subsection{X-ray spectral analysis} 
 
\begin{figure*}
    \centering
    \includegraphics[height=10.5cm,width=8.5cm,angle=-90]{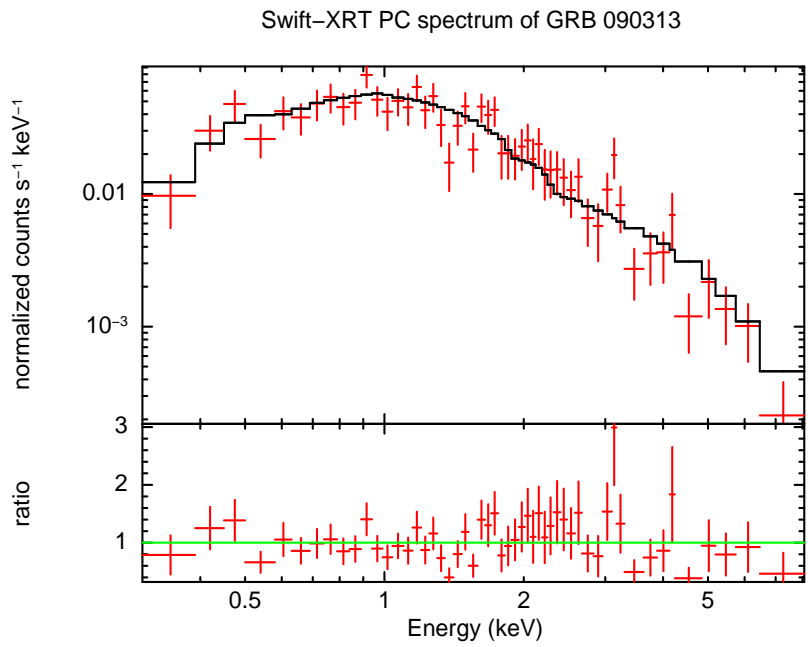} 
    \caption{The X-ray spectrum in the time interval [T0+26779, T0+46710] s is well fitted by a simple absorbed power law (Evans et al. 2007).}
    \label{figXrayspec}
 \end{figure*}

The X-ray spectrum (Fig. \ref{figXrayspec}; from the {\it Swift}-XRT repository, Evans et al. 2007) can be fitted by an absorbed simple power law with a photon index $\Gamma_X = 2.14^{+0.12}_{-0.14}$ and an absorbing column density $N_{\rm H} = (2.99^{+0.77}_{-0.71}) \times 10^{22}$ cm$^{-2}$, in excess of the Galactic value of $2.1 \times 10^{20}$ cm$^{-2}$.

\subsection{Spectral energy distribution} 
 
\begin{figure*}
    \centering
    \includegraphics[height=10.5cm,width=8.5cm,angle=-90]{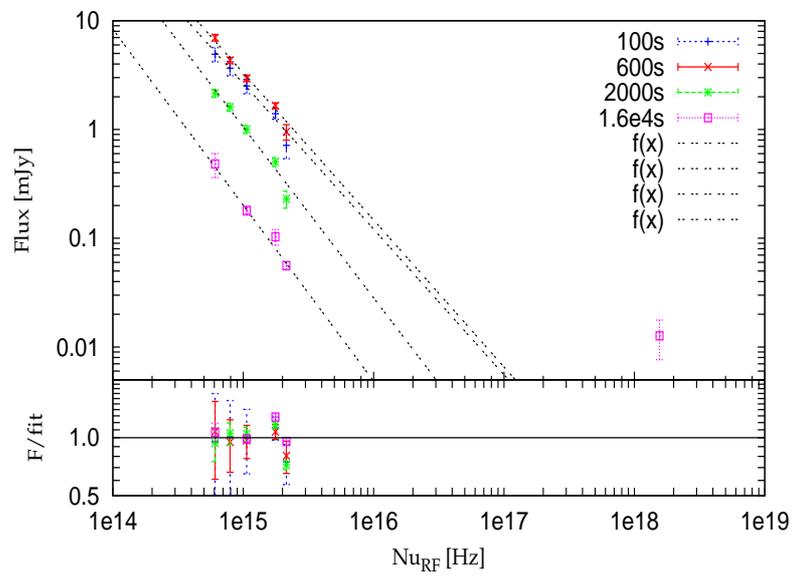} 
    \caption{Multi epoch spectral energy distribution. Times are in the rest frame of the GRB.}
    \label{figSED}
 \end{figure*}

 From our data and others published in GCNs we 
 estimate the flux for the infrared ($JHK$) and optical ($i'r'$) filters
 at four different epochs (corresponding to T0+100 s, T0+600 s,
 T0+2$\times 10^{3}$ s and T0+1.6$\times 10^{4}$ s in the rest frame of
 the burst). At the redshift of the burst (z=3.374) the wavelength of the Lyman-alpha break
 (121.6 nm) is redshifted to 532 nm, that corresponds roughly to the
 central peak wavelength of the $V$ filter. However also the tail of the
 $R$ filter could be affected by the absorption and for that reason we
 decided to perform the fit of the optical spectral energy distribution
 only up to 2$\times 10^{15}$ Hz. The results of the fit are shown in
 Fig. $\ref{figSED}$ and reported in Table \ref{tabsed}. The afterglow
 of GRB 090313 did not display any spectral evolution before and after
 the peak in the light curve. Only a slight and insignificant change of
 the spectral parameter $\beta_{\rm O}$ is recorded around 3ks (observed
 frame) after the break. For this reason we built a composite
 optical/infrared light curve (fixing the value of $\beta_{\rm O} = 1.2$) 
using rigid shifts for each filter  to report all the fluxes relative to the SDSS-$i$ band. 
 
 \section{Discussion} 
 
Here we examine the properties of 19 GRBs including GRB 090313 that
exhibit a single-peaked optical light curve. Those are all the GRBs with published data that 
show a clear rise and fall of their optical light curves. The observed and derived
properties of the sample are given in Table \ref{tabprop}. In this table we 
report the parameters of the optical peak ($\alpha_{\rm rise}$, $\alpha_{\rm decay}$, 
t$_{\rm peak}$ and $F_p$), together with the X-ray decay index
($\alpha_{\rm X}$) in the post optical peak phase \footnote{~the value of $\alpha_{\rm X}$ is taken from the
literature or from the XRT light curve repository (Evans et al. 2007).}, the duration (T$_{90}$), 
redshift ({\it z}), initial Lorentz factor $\Gamma$
and isotropic energy (E$_{\rm iso}$) for each burst. We have assumed that the 
optical peak time represents the fireball deceleration time. Following equation 1 in Molinari et al. 2007, the initial Lorentz factor 
of  GRB 090313 is give by 
\begin{equation}
\Gamma \approx 80~n^{-1/8}  
\left(\frac{E_{\rm iso}}{3.2 \times 10^{52}~{\rm erg}}\right)^{1/8}
\left(\frac{1+z}{4.375}\right)^{3/8}
\left(\frac{t_{\rm peak}}{1060~{\rm s}}\right)^{-3/8}
\end{equation}
where $n$ is the ambient density in protons/cm$^3$. For all the bursts in Table \ref{tabprop} 
the ISM environment is favored in literature (i.e. Klotz et al 2008, Rykoff et al. 2009, Oates et 
al. 2009, Melandri et al. 2009, Greiner et al. 2009); only GRB 080330 is better explained by a 
wind-like medium (Guidorzi et al. 2009). For the wind medium $\rho=A R^{-2}$, the equation 1 
is replaced by $\Gamma \sim 25 ~ (A/5 \times 10^{11} ~{\rm g~cm^{-1}})^{-1/4} (E/3.2 \times 10^{52}~{\rm erg})^{1/4} 
[(1+z)/4.375]^{1/4} (t_{\rm peak}/1060 ~{\rm s})^{-1/4}$.
 
It is well accepted that the X-ray temporal decay of the majority of GRB
afterglow can be described by a canonical light curve, where the initial
X-ray emission (steep decay) is consistent with the tail of the gamma-ray 
emission, followed by a shallow phase that leads into a power-law decay 
phase (Nousek et al. 2006, Zhang et al. 2006, Tagliaferri et al. 2005,
O'Brien et al. 2006). In our sample also, no peaks are detected in 
the X-ray light curves, all the X-ray light curves monotonically decay
from the beginning of the X-ray observations except X-ray flares. 
It is known that about 50$\%$ of GRBs show flaring activities on top of the 
canonical light curve. The narrow structure $\Delta t/ t <1$ indicates that it 
originates from a physically distinct emitting region (e.g. late internal shocks).
X-ray observations started before an optical peak for GRB 990123,  GRB 050730, 
GRB 050820A, GRB 060418, GRB 060605, GRB 060607A, GRB 060904B, 
GRB 070419A, GRB 074020, GRB 071031, XRF 080330 and GRB 080810, while 
it started after an optical peak for GRB 061007, GRB 080603A, GRB 080129, 
GRB 080710 and GRB 090313. We have no X-ray observations for XRF 020903 
and XRF 030418. If an optical peak is due to the deceleration of a fireball, 
X-ray emission from external shocks also should peak simultaneously.
The tail of the prompt emission or a different emission component 
might mask the X-ray peak. 

Four events: GRB 060418, GRB 060605, GRB 060607A and GRB 060904B
show X-ray flares around an optical peak, we tested whether the observed
optical peaks could be explained by the flare emission alone by
extrapolating the peak flux of the X-ray flare to the optical band
assuming a spectral index between the two bands of $\beta \sim 1$. In
all cases, the contribution of the X-ray flare to the optical light curve was
significantly lower than that observed, ruling out a flare origin for the
optical peaks.

  \subsection{The origin of the optical peak}
 
Recent results on the naked eye optical flash from GRB 080319B (Racusin
et al. 2008; Bloom et al. 2009), where the observed optical peak coincided 
in time with the prompt gamma-ray emission, provided motivation to consider 
that the prompt gamma-ray emission is Inverse Compton (IC) of the optical 
flash. The dominance of IC cooling could lead to the lack of prompt optical 
flashes.\footnote{The full discussion on IC cooling effects (e.g. Nakar, Ando $\&$ 
Sari 2009) is beyond the scope of this paper. We here give a rough estimate on 
how much $\epsilon_B$ would be necessary to suppress an optical flash. We 
assume that the typical frequency of the reverse shock is in the optical band, 
and that the shock emissions are in the fast cooling regime. If the IC cooling 
is not important, the luminosities would peak at the typical synchrotron 
frequencies, and the luminosities would be comparable at the onset of afterglow. 
The flux ratio is about $\Gamma$ in the optical band (Kobayashi $\&$ Zhang 2003). 
If the IC cooling is the dominant cooling mechanism of the electrons in the shock 
regions, the bulk of the shock energy is radiated in high energy radiation 
(possibly the 1st scattering component for the forward shock and the 2nd 
scattering component for the reverse shock). The optical flux ratio could be 
reduced roughly by a factor of $(\epsilon_e/\epsilon_B)^{1/6}$ (Kobayashi 
et al. 2007). A very small $\epsilon_B \sim \epsilon_e/\Gamma^6$ is required 
to explain the lack of optical flashes.  In the slow cooling regime, the Compton 
parameter is smaller for a given ratio $\epsilon_e/\epsilon_B$, the required 
$\epsilon_B$ could be even smaller.} However, the basic problem of such IC 
model is that if the low-energy seed emission is in the optical, while the 
observed soft gamma-ray spectrum is the first IC component, then second 
IC scattering would create a TeV component. The second IC component in the 
TeV range should carry much more energy than the soft gamma-ray components. 
This could cause an energy crisis problem, possibly violating upper limits from 
EGRET and Fermi (Piran et al. 2009).    
 
 Rykoff et al. (2004) suggested a model in which single-peaked light
 curves are caused by GRB radiation emerging from a wind medium
 surrounding a massive progenitor. This model suggests that the rise of the
 afterglow observed in the optical band can be ascribed to extinction
 and the emission can be modeled with an attenuated power-law. A
 consequence of this model is that at very early times some afterglows
 will rise very steeply and the extinction observed in the optical band
 should be much greater that in the infrared band. As shown in
 Fig.\ref{figp1} we see a very steep rise only for GRB 061007, however
 for this burst as for the other bursts on that figure, we do not have data to 
 model the peak in the infrared band. If we fit the afterglow peak of GRB 090313 
 with an attenuated power-law function (equation 1 in Rykoff et al. 2004) we 
 find values of the decay index and the attenuation time scale ($\alpha$ = 1.15 
 $\pm$ 0.03 and $\beta_{\rm t}$ = 1097 $\pm$ 117 s) consistent with the decay 
 index $\alpha$ obtained in section 3.2. With this $\beta_{\rm t}$ we derive 
 a mass loss rate ($\sim 10^{-3}$~M$_{\odot}$~yr$^{-1}$) which is slightly 
 higher than what is usually suggested for GRB progenitors. The Lorentz factor 
 that we assumed for this estimate is obtained from the peak time based on the 
 wind model;  $\Gamma$ based on the ISM model is higher and it would results 
 in a higher mass loss rate. This is a similar result to the one found by Rykoff et al. 
 for GRB 030418. As the majority of the GRBs in our sample rise slowly or with 
 comparable $\alpha_{\rm rise}$ with respect GRB 090313 this will imply a higher 
 mass loss rate for all those bursts. This model will be further tested with future 
 simultaneous optical/IR light curves obtained at early time.

If the observed peak is due to the passage of the typical frequency of
the forward shock through the optical band, we would expect much slower
rise ($\alpha_{\rm rise} \sim$ -0.5) and strong color evolution around
the peak. These are not consistent with GRB 090313 observations 
($\alpha_{\rm rise} \sim$ -1.7 and no color evolution). If the optical peak 
is due to the deceleration of a fireball,
the typical frequency of the forward shock $\nu_{\rm m,fs}$ should be
below the optical band at the onset, otherwise, the forward shock
emission slowly rises until the typical frequency crosses the optical
band. Actually when this condition: 
$\nu_{\rm m, fs} (t_{\rm peak }) < \nu_{\rm optical}$ is satisfied, 
the forward and reverse shock emission peak at the same time, and 
produce a single peak (Mundell et al. 2007a).  We here consider such a
low-frequency model in detail. 

The onset of the afterglow is expected to occur immediately after the prompt
emission if the reverse shock is in the thick shell regime, while there
should be a gap between the prompt gamma-ray emission and the onset if
the reverse shock is in the thin shell regime (Sari 1997). At the onset
of afterglow, the forward and reverse shock emission rise as $F\propto
t^{3}$ and $t^{3p-3/2}$, respectively in the thin shell case, 
while they are as shallower as $t^{(3-p)/2}$ and $t^{1/2}$ for the thick
shell case. If the two emission components are comparable at the onset,
the rising index could be determined by the shallower component. The rising
index is expected to be $t^3$ for the thin shell case, and $t^{1/2}$ or shallower for
the thick shell case. 

As we will discuss, most optical afterglows are classified into 
the thin shell case. The fireball deceleration time is given by
$t_{peak} \sim 90~(1+z) E_{52}^{1/3}n^{-1/3}\Gamma_2^{-8/3}$ s
where we have scaled parameters as $E_{52}=E_{\rm iso}/10^{52}$ergs and 
$\Gamma_2=\Gamma/100$. At the peak time $t_{\rm peak}$, the cooling
frequency and the typical frequencies of the forward and reverse shock
emission are given (Sari et al. 1998, Kobayashi $\&$ Zhang 2003) by  
\begin{eqnarray}
\nu_{c}         &\sim&  2.6\times 10^{18} 
~(1+z)^{-1}\epsilon_{B,-3}^{-3/2} E_{52}^{-2/3}n^{-5/6}\Gamma_2^{4/3} ~{\rm Hz},\\
\nu_{\rm m, fs} &\sim&  5.4 \times 10^{13} 
~(1+z)^{-1}\epsilon_{e,-2}^{2} \epsilon_{B,-3}^{1/2} 
n^{1/2}\Gamma_2^{4}  ~{\rm Hz},\\
\nu_{\rm m, rs} &\sim&  5.4 \times 10^{9} 
~(1+z)^{-1}\epsilon_{e,-2}^{2} \epsilon_{B,-3}^{1/2} 
n^{1/2}\Gamma_2^{2}  ~{\rm Hz},
\end{eqnarray}
where $\epsilon_{e,-2}=\epsilon_{e}/10^{-2}$ and 
$\epsilon_{B,-3}=\epsilon_{B}/10^{-3}$. For plausible parameters,
the typical frequency of the forward shock is actually below optical band 
and the both shock emission is in the slow cooling regime. 
 
The low typical frequencies provide an upper limit to the microscopic 
parameter $\epsilon_e$. Requiring that the typical frequency of the 
forward shock is below the optical band at the onset of afterglow, we obtain
\begin{equation}
\epsilon_e \leq 0.30 ~
\left(\frac{\epsilon_B}{0.003}\right)^{-1/4} ~
(1+z)^{-1/4}~ 
\left(\frac{t_{\rm peak}}{30 ~{\rm min}}\right)^{3/4} ~  
\left(\frac{E_{\rm iso}}{10^{52} ~{\rm ergs}}\right)^{-1/4}
\left(\frac{\nu_{\rm opt}}{10^{15} ~{\rm Hz}}\right)^{1/2} 
\end{equation}
The estimated values for the upper limit of  $\epsilon_e$ for the GRBs in our 
sample are reported in table \ref{tabprop}. The spread of the values of $\epsilon_B$ is large (from $\sim 10^{-4}$ to $\sim 10^{-1}$) and this could still be a significant uncertainty in the upper limits estimates, even if $\epsilon_e$ do not strongly depend from that parameter. GRB 020903 does not constrain 
$\epsilon_e$ well, the typical upper limit is $\sim 0.08$, consistent with values 
from later afterglow modeling (e.g. Panaitescu $\&$ Kumar 2002).

A highly magnetized fireball is another possibility to explain the lack of optical 
flashes.\footnote{The reverse shock emission might be suppressed for high 
magnetization: $\sigma = B^2/4\pi \rho c^2 \sim 0.1$ or larger where $B$ and $\rho$ 
are the rest-frame magnetic field strength and density, respectively (Mimica et al. 2009). 
Assuming a mildly relativistic reverse shock, the critical magnetization could 
correspond to $\epsilon_B \sim 0.1$.} However, Granot et al. (2010) recently argued 
that in the thin shell case the magnetization of the GRB outflow at the deceleration 
time is not high enough to suppress the reverse shock. 
Most events in our sample are classified into the thin shell case. Even if 
the reverse shock is suppressed by high magnetization, the same condition 
$\nu_{\rm m,fs}(t_{\rm peak}) < \nu_{\rm optical}$ could be required to
avoid slowly rising forward shock emission after the onset of afterglow.

\subsection{Reverse and Forward Shocks: Relative Contributions}

\begin{figure*}
    \centering
    \includegraphics[height=6.5cm,width=7.0cm]{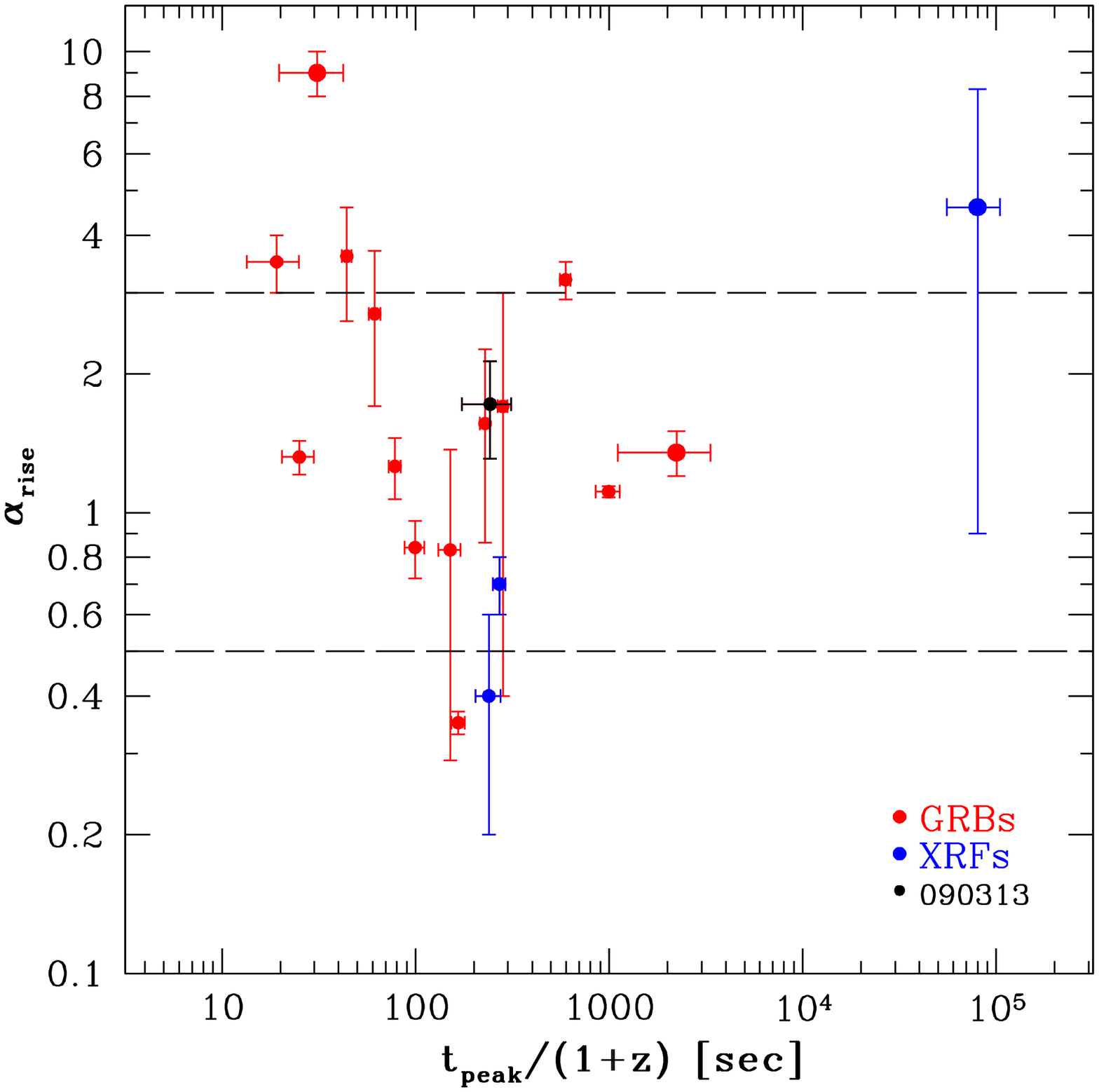} 
    \includegraphics[height=6.5cm,width=7.0cm]{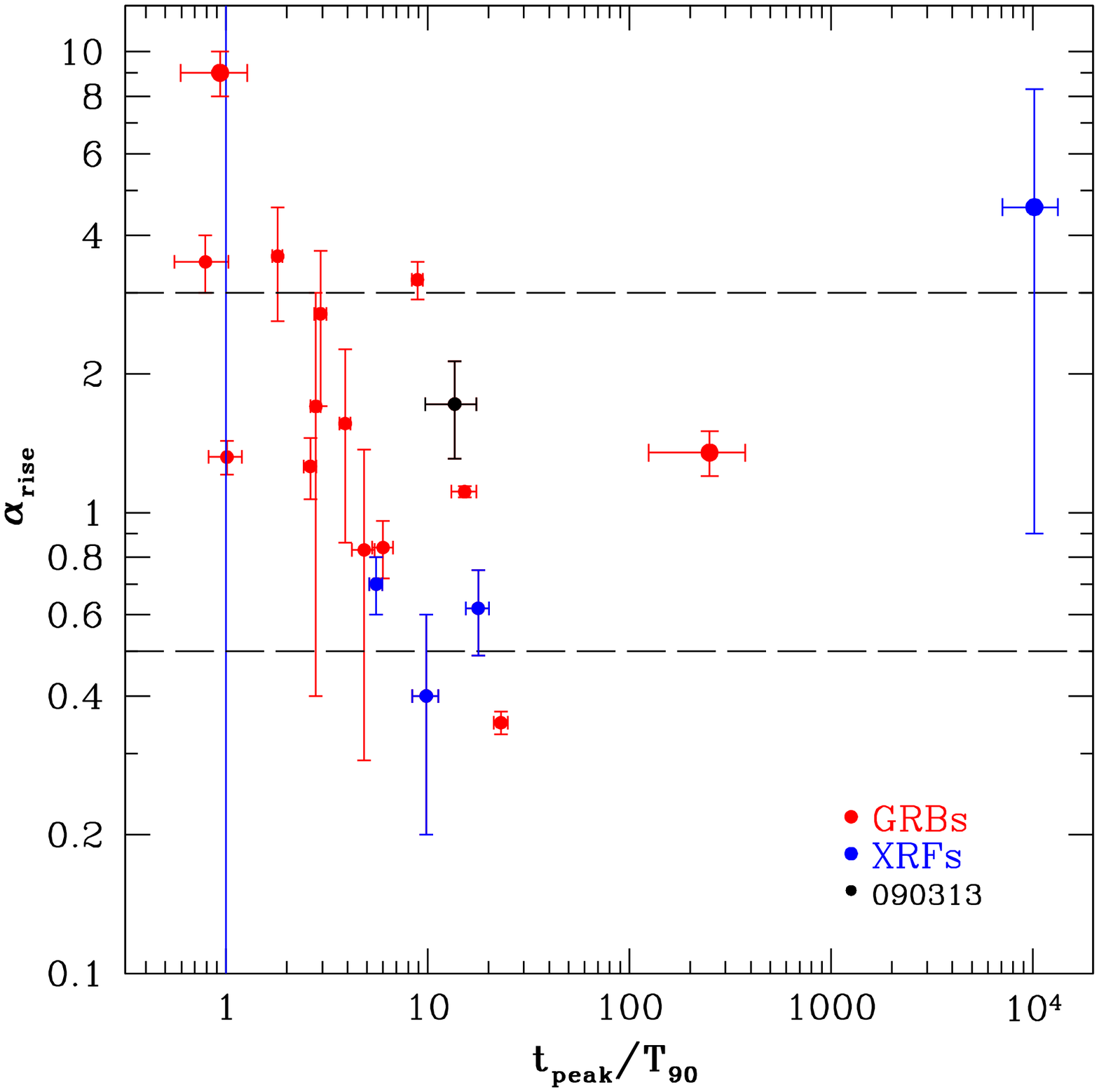} 
    \caption{Left: rise index ($\alpha_{\rm rise}$) vs time of the peak in the GRB rest frame. Right: rise index ($\alpha_{\rm rise}$) vs t$_{\rm peak}$/T$_{90}$. Reverse shocks can be classified into two classes: 1) t$_{\rm peak} >$ T$_{90}$, thin shell case, sharp rise; 2) t$_{\rm peak} \sim$ T$_{90}$, thick shell case, slow rise. The vertical line shows where t$_{\rm peak}$ = T$_{90}$. Dashed horizontal lines represent the asymptotic values for the two cases (see text for details). The absolute values of $\alpha_{\rm rise}$ are plotted.}
   \label{figp1}
 \end{figure*}

In Fig. \ref{figp1} we plot the light curve rise index ($\alpha_{\rm
rise}$) against the time of the peak in the GRB rest frame (left panel)
and the ratio $t_{\rm peak}$/T$_{90}$ (right panel). In a recent work 
Panaitescu $\&$ Vestrand (2008) classified the optical light curves into
'fast-rising with an early peak' and 'slow-rising with a late peak'. In our sample
an apparent weak anti-correlation can be seen between $\alpha_{\rm rise}$ 
and $t_{\rm peak}$; however the significance is very low ($\sim 12\%$) not 
allowing any firm conclusion about the existence of this anti-correlation.
A simple fireball model predicts that the dynamics of a fireball is classified 
into two cases: (1) thin shell fireballs (t$_{\rm peak} >$ T$_{90}$) produce a sharp 
peak with rising index $\alpha_{\rm rise} \sim -3$; (2) thick shell fireballs ($t_{\rm peak} 
\sim $T$_{90}$) have a wider peak with $\alpha_{\rm rise} \sim -1/2$ (where this value is a limit and the rise could be much shallower). 
As shown in Fig. \ref{figp1} (right panel), most GRBs in the sample are classified
into the thin shell case, and the rising indexes are consistent with the
simple model or shallower. The simple reverse shock model assumes a
homogeneous fireball. However, as internal shock process requires, the
initial fireball could be highly irregular. The complex structure of
shell or energy injection in the post-prompt  phase could make
the rising index shallower.

In Fig. \ref{figp1}, GRB 061007 stands out as a notable exception with
Rykoff et al. (2009) quoting a peculiarly steep rising index
($\alpha_{\rm rise} \sim$ -9). Mundell et al. (2007a)  
showed that the afterglow is detected from gamma to optical wavelength, beginning 
during the prompt emission as early as 70~s post-trigger. The softening of the 
gamma-ray spectral index after 70~s further confirms the afterglow onset at this time  
(Mundell et al. 2007a; Rykoff et al. 2009). The gamma ray light curve is dominated by a 
multi-peaked flare between T=20 and 70~s, coincident with the steepest rising part of 
the optical light curve and possible double optical peak. If the optical emission during 
these prompt gamma-ray flares comprises a rising afterglow component with a 
contemporaneous prompt (flaring) component superimposed, the underlying
afterglow rising index would be much shallower than the observed value.
In our small sample, the optical afterglow of XRFs tend to rise slowly
with a late peak. If we ignore XRFs and the peculiar case of GRB 061007
in fig \ref{figp1}, the anti-correlation between the rising index and
peak time is very weak or it might not exist.

GRB 990123 has a clear reverse shock component in early optical afterglow. Our low-frequency 
model is not suitable to discuss this event, because it is considered to explain the lack of optical 
flash. On the other hand, the simple reverse shock model still predicts that the rising index is $\sim 1/2$ 
 for the reverse shock dominant thick shell case. We plot GRB 990123 in Fig. \ref{figp1} also to test 
 the simple model. The discrepancy might be due to the irregularity of the fireball.

 An interesting comparison can be done with the peaks detected in the
 high energy band by the {\it Fermi}/LAT. Ghisellini et
 al. (2010) studied the emission observed at energies $>0.1$ GeV of 
 11 GRBs detected by the Fermi. They argue that the observed high 
 energy flux can be interpreted as afterglow emission shortly following
 the start of the prompt emission. Most events show the onset of 
 afterglow during the prompt gamma-ray phase. This is quite a contrast 
 to what we have seen in our sample. The reason for this difference 
 might be that Fermi events tend to have very high Lorentz factors, 
 which allow to emit high energy photons without pair attenuation, 
 and the events are classified into the thick shell. The peak time 
 should be early and comparable to the duration of the prompt emission. 
 On the other hand, our sample (early optical observations) might be biased
 towards the thin shell case, because early peaks are technically
 difficult to catch.   

   \subsection{Radio Afterglow Modeling}
 
In the low frequency model, the characteristics of an optical peak: the peak time 
$t_{\rm peak}$ and the peak flux $F_p$ can be used to predict the behavior of 
early radio afterglow. At the onset of afterglow (peak time), the typical frequencies 
and spectral peaks of reverse and forward shock are related as $\nu_{\rm m,rs}\sim\Gamma^{-2}\nu_{\rm m,fs}$ 
and $F_{\rm max,rs} \sim \Gamma F_{\rm max,fs}$, respectively (Kobayashi
\& Zhang 2003). Note that $F_p$ is a peak in the time domain, while $F_{max}$ is 
a peak in the spectral domain. To produce bright forward shock emission, 
$\nu_{\rm m, fs}$ should be close to optical band and  we get $\nu_{\rm m,rs}\sim \Gamma^{-2}\nu_{\rm
opt}$ and $F_{\rm max,rs} \sim \Gamma F_p$. After the original fireball deceleration, 
the typical frequency and spectral peak behave as $\nu_{\rm m,rs} \sim t^{-3/2}$ 
and $F_{\rm max,rs} \sim t^{-1}$. The typical frequency comes to the radio band at 
$t\sim \chi^{2/3}(\nu_{\rm opt}/\nu_{\rm radio})^{2/3}
\Gamma^{-4/3} t_{\rm peak} \sim 2 \times10^3 \Gamma^{-4/3} t_{\rm peak}$ 
and the flux at that time is $F\sim \chi^{-(3p+1)/6} (\nu_{\rm opt}/\nu_{\rm radio})^{-2/3}\Gamma^{7/3}F_p
\sim 5 \times10^{-4}\Gamma^{7/3}F_p$ where $\chi=\nu_{\rm m,fs}/\nu_{\rm opt} < 1$ 
is a correction factor when $\nu_{\rm m,fs}$ is well below 
the optical band and $F_p = \chi^{(p-1)/2}F_{\rm max,fs}$, in principle, $\chi$ could be determined from radio 
observations, $\Gamma$ is estimated from the peak time $t_{\rm
peak}$ as shown in table \ref{tabprop}. In eq (5), the upper limit corresponds 
to the case of $\chi = 1$. If $\chi$ is obtained from radio observations, the 
right-hand side of the inequality with a correction factor of $\chi^{1/2}$ 
gives the value of $\epsilon_e$.

\begin{figure*}
    \centering
    \includegraphics[height=6.5cm,width=7.0cm]{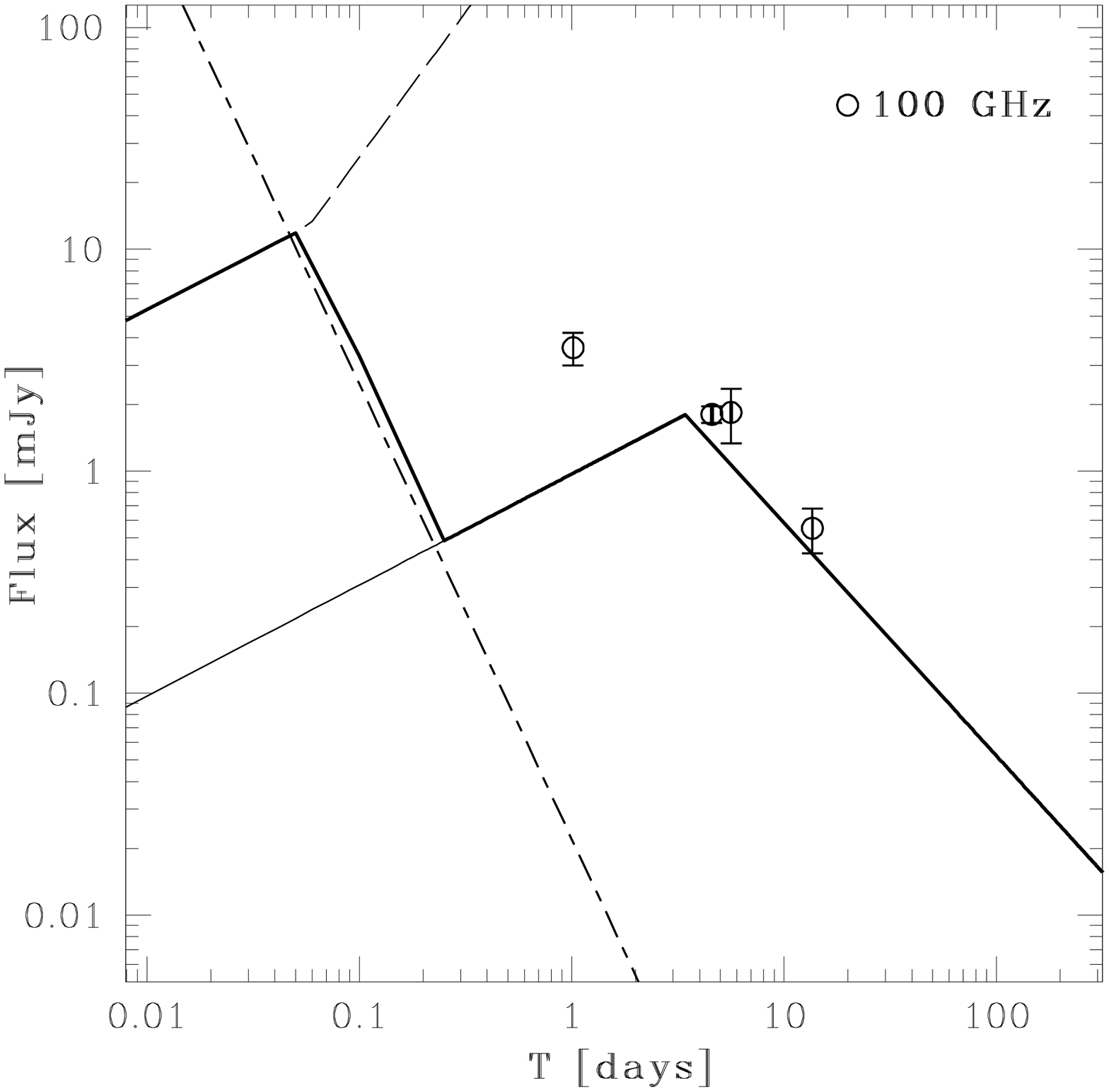} 
    \includegraphics[height=6.5cm,width=7.0cm]{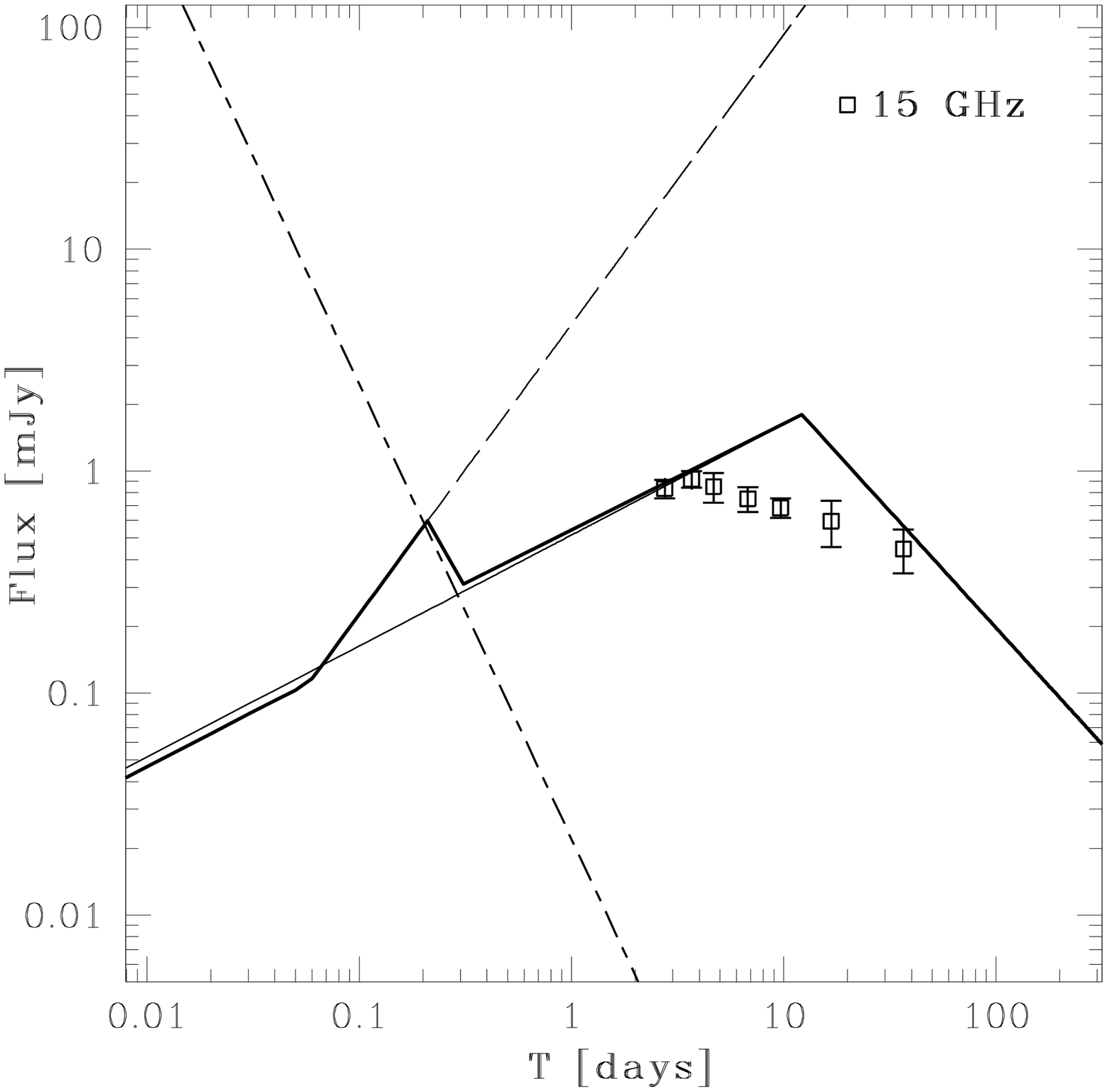} 
    \caption{Expected light curves at  100 GHz (left) and 15 GHz (right). The reverse shock evolution is shown as a dot-dashed line, the self-absorption curve as a dashed line and the forward shock evolution as a solid thin line. The thick line represent the expected light curve for GRB 090313. Circle and square points represent observed data in the mm (100 GHz) and radio (15 GHz) band respectively.}
    \label{figradio}
 \end{figure*}

At low frequencies and early times, self-absorption takes an important
role and significantly reduces the flux. A simple estimate of the maximal 
flux is the emission from the black body with the reverse shock
temperature (Sari \& Piran 1999; Kobayashi \& Sari 2000). The black body
flux at the peak time is 
\begin{equation}
F_{\nu,BB} \sim \pi (1+z)\nu^2 \epsilon_e m_p \Gamma \left(\frac{R_\bot}{D_L}\right)^2
\end{equation}
where $R_\bot \sim 2 \Gamma ct_{\rm peak}$ is the observed size of the
fireball. This limit initially increases as $\sim t^{1/2}$, and then steepen 
as $\sim t^{5/4}$ after $\nu_{\rm m,rs}$ crosses the observation frequency $\nu$.
In Fig. \ref{figradio} the dashed lines indicate the black body flux limit. Once 
the reverse shock emission becomes dimmer than the limit, the flux decays 
as $\sim t^{-(3p+1)/4}$. The combination of the increasing limit and decaying
flux shapes ``radio flare'' (Kulkarni et al. 1999). 
The forward shock emission (thin solid)
evolves as t$^{1/2}$ before the passage of $\nu_{\rm m,fs}$ through the
radio band, and then decays as  $t^{-3(p-1)/4}$. The forward shock peak
$F\sim \chi^{-(p-1)/2} F_p$ should happen around $t\sim \chi^{2/3}
(\nu_{\rm opt}/\nu_{\rm radio})^{2/3} t_{\rm peak} \sim 2 \times10^3 t_{\rm peak}$ s.

In the case of GRB 090313, assuming $\chi \sim 1$, the forward shock
peaks in the optical band $\nu_{\rm m,fs}$ $\sim$ $4.6 \times 10^{14}$ Hz with a flux density
F$_{\rm max,fs} \sim$ 2 mJy and a peak time corresponding to
$\Gamma \sim$100; therefore the reverse-shock peak flux at this
time occurs at $\nu_{\rm m,rs}$ (t$_{\rm peak}$) $\sim$ 46 $\times
10^{9}$ Hz  and is  F$_{\rm max,rs} \sim$ 200 mJy.  Correcting for
synchrotron self-absorption, results in an observable flux density of
$\sim$ 4 mJy after 2.4 hours.  After the deceleration time the
reverse-shock emission in the radio band decays as
$\sim t^{-2}$ (dot-dashed line
Fig. \ref{figradio}) and the emission at 1 day is about $\sim$ 20
$\mu$Jy for GRB~090313.  

For GRB 090313, the forward shock emission is expected to peak in the
radio band around 12 days after the burst (assuming $\nu_{radio} = 
1.5 \times 10^{10}$ Hz), with peak flux of 2 mJy 
(solid line Fig. \ref{figradio}).  Taking all these factors into
account,  the resultant expected light curve of the radio afterglow of
GRB 090313 is shown in Fig. \ref{figradio}. The expected 15 GHz
and 100 GHz light curves (thick lines) are reasonably consistent with the
observations. The deviation of the 15 GHz estimates from the 
observations might be partially due to a simplified synchrotron spectrum 
which is described by a broken power law. The deviation of the 100 GHz 
point around 1 day is much more apparent and might be due to an additional 
emission component (e.g. late time central engine activity).
Since a realistic synchrotron spectrum is rounded at the break
frequencies, a more accurate estimate should give a light curve 
rounded at the peak time. However, if this is the case, our simple model
further underestimates the 100 GHz flux. It is interesting that 
the 15 GHz flux decays very slowly up to few ten days while X-ray
afterglow displayed a steep decay around $\sim$ 1 day, as shown in
Section 3.2. This might indicate different origins (e.g. emission regions) for 
the two; the $\delta\alpha = \alpha_{2} - \alpha_{radio} > 2$ is indeed too 
large to be explained assuming that the cooling frequency lies at the 
X-ray frequencies at that time. 

\begin{figure*}
    \centering
    \includegraphics[height=6.5cm,width=7.0cm]{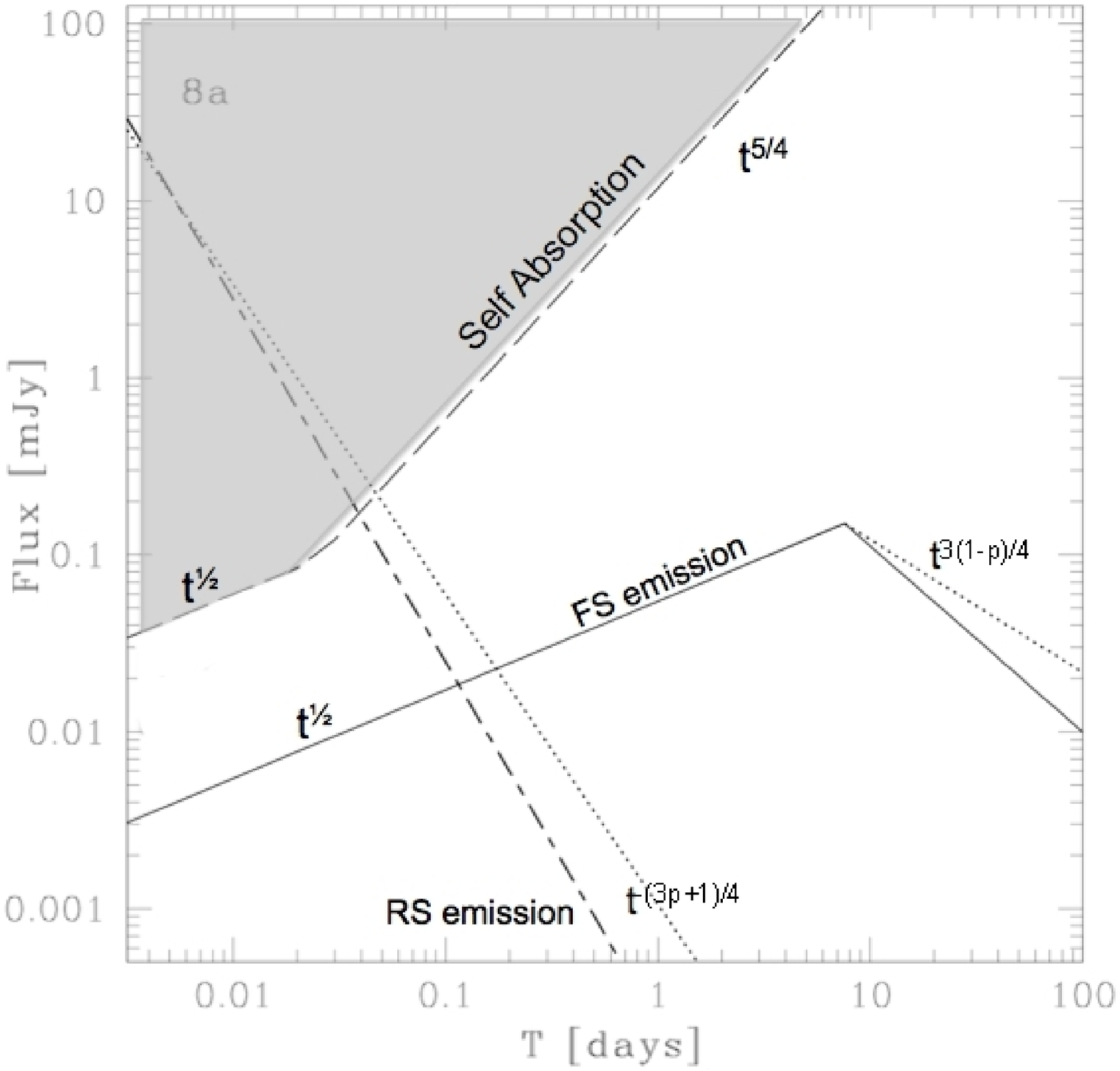} 
    \includegraphics[height=6.7cm,width=7.0cm]{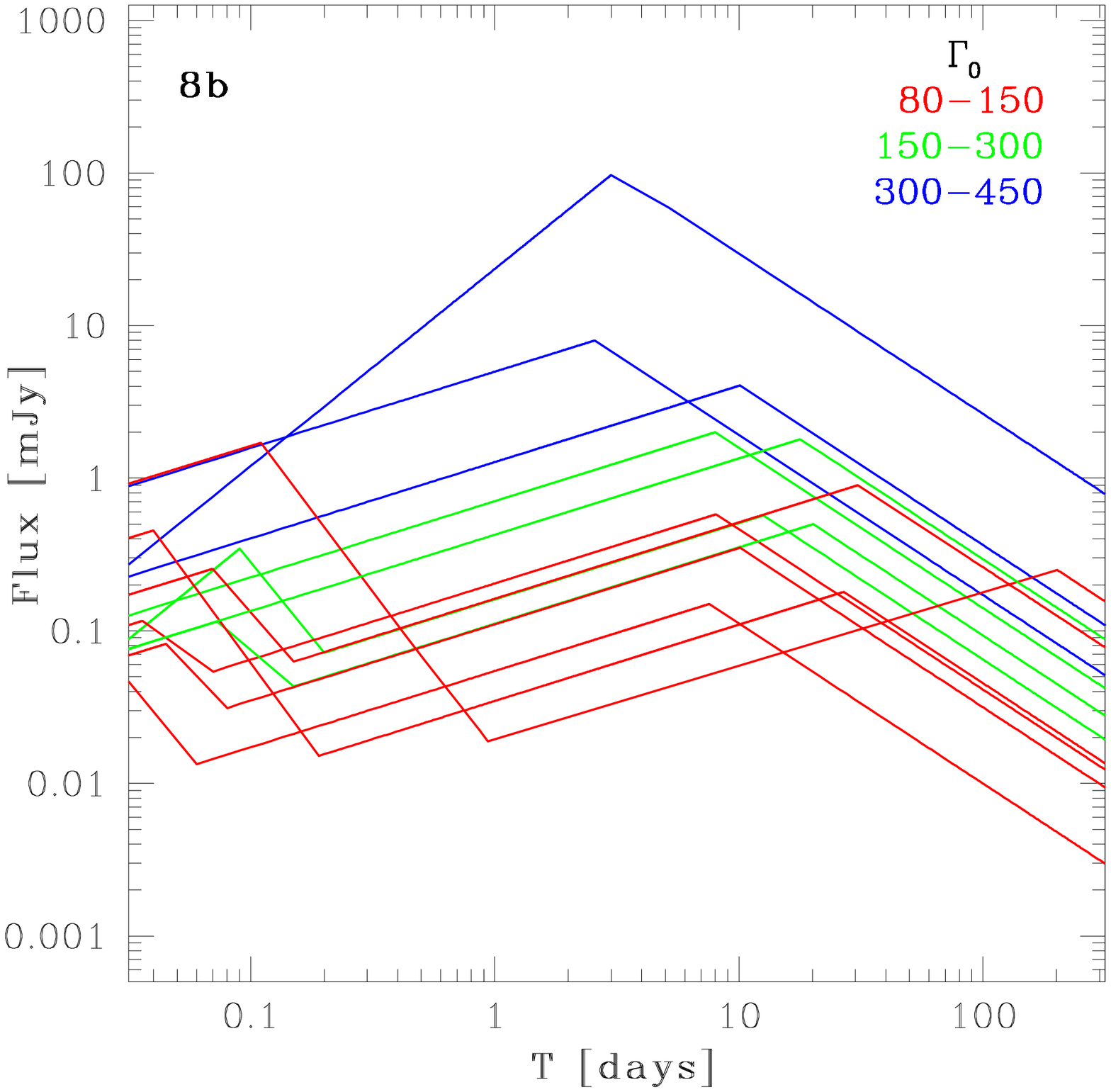} 
    \caption{Schematic showing general dependencies for different components (a); predicted light curves in the radio band, $\nu$ = $\nu_{\rm VLA}$ = 8.5 GHz (b) for the sample GRBs, derived from their observed optical properties listed in Table \ref{tabprop}. The brightest blue curve represent the peculiar high energetic case of GRB 061007 for which a very bright radio afterglow is expected, see text for more details.}
    \label{figradio2}
 \end{figure*}

In Fig \ref{figradio2}, we show radio light curves expected for 
our sample, which are evaluated by using early optical observations. 
GRB 990123, XRF 020903, GRB 030418, GRB 060607A, GRB 070420 
and GRB 080810 are excluded in the radio afterglow estimates. Since 
GRB 990123 clearly shows a reverse shock component in the early 
afterglow, it is not consistent with our model assumption. For the other 
five events, the optical peak time or peak flux was not well constrained. 
In future, we should be able to estimate radio afterglow light curves in real 
time as soon as a single peaked optical light curve is detected. Depending 
on the Lorentz factor at the time of the peak and on the energetics of the
burst, the shape of the radio will slightly change, displaying an early
peak/flash at $\sim$ 0.1 days and later on the peak of the forward shock
in the radio band peaking at about 2-10 days after the burst. Diffractive 
scintillation might make the detection of radio flares difficult if the amplitude 
of flares are order of unit. In cases similar to GRB~061007, in which the 
optical forward and radio reverse shocks peak at early time and the 
forward shock flux is large, the radio peak due to the the passage of 
the forward shock typical frequency is expected to be very bright. Liang et al. (2010)
suggest a correlation, such that $\Gamma \propto E_{\rm iso}^{2/7}$,
therefore Fig. \ref{figradio2} can also be viewed in terms of increasing
E$_{\rm iso}$. The scatter in this correlation, however results in an
over-prediction of the the initial Lorentz factor for GRB 090313 ($\Gamma \sim$ 130) compared
with the value calculated directly from the light curves.

\section{Conclusions} 
 
 We have analysed multiwavelength observations of GRB~090313 and  
      similar 18 GRBs which exhibit a single-peaked optical light curve. We have 
      compared prompt and afterglow properties to test the standard
      fireball model with amended microphysics parameters. The goal of the
      study was to understand the origin of single optical peaks in
      afterglow light curves and to explain the surprising lack of bright
      optical flashes from reverse shocks that were predicted from 
      the standard fireball model.
      Within this amended standard model, which
      we term the {\em low-frequency model}, a single peak in the early
      time optical light curve is produced when the typical synchrotron
      frequencies of shock emission lie below the optical band.
      We have shown that this condition is satisfied with plausible
      microphysics parameter $\epsilon_e$; the single peak
      consists of forward and reverse shock emission components, the
      peak time represents the initial deceleration of the fireball at
      the onset of the afterglow and the reverse shock emits most
      photons at frequencies below the optical band. We find that:

 \begin{itemize}

\item In the case of GRB~090313, no spectral evolution was observed at
      the time of the optical peak,
      the peak is considered to represent the onset of the GRB
      afterglow (or fireball deceleration) and the initial Lorentz
      factor of the ejecta was derived $\Gamma \sim 80$. The
      Lorentz factors that were similarly derived for the other GRBs and
      XRFs in the sample cover a wide range $40 < \Gamma <
      450$.
      
\item The rising indexes of most optical light-curves
      are consistent or shallower than the value of $F\sim t^3$
      expected in the standard model. Although a simple reverse 
      shock model assumes a homogeneous fireball, the internal shock
      model requires a highly irregular fireball. At the end of 
      the prompt gamma-ray phase, the fireball might still have 
      an irregular structure. The irregularity in the density 
      distribution or energy injection in post-prompt gamma-ray phase 
      could make the rising index shallower than the expected value. 
      In the small sample, the optical 
      afterglow of XRFs tend to rise slowly with a late peak. 
            
\item We constrained the value of $\epsilon_e$ for the single-peak
      events, found  an average value of $< 0.08$ for the whole
      sample. The values derived from early time light curve properties
      are consistent with published values derived from late-time
      afterglow modeling. However the large spread of values for $\epsilon_B$ could
      affect the estimates of the upper limit for $\epsilon_e$.

\item Using the observed optical properties for our sample of GRBs, we
       predicted the radio afterglow light curves for the low-frequency
       model. Synchrotron self-absorption is important at early times in
       shaping the radio light curve and masking the reverse shock
       emission. This could result in an early detectable peak around
       $\sim$ 0.1 days, though prompt radio observations might be 
       challenging. The forward shock peaks later around 2-10 days
       after the burst.  It is important to note that high energies and
       Lorentz factors (as in the case of GRB 061007) could produce
       bright optical and radio afterglows. We demonstrate the effectiveness 
       of this method in the case of GRB 090313. This is important for future
       observations of GRBs afterglow in the radio band with new
       facilities such as the EVLA, ALMA and LOFAR. The latter will have a very large 
       field of view, and prompt radio observations could  be possible. However, 
       LOFAR will operate at low frequencies (below 250 MHz) and since synchrotron 
       self-absorption limit F$_{\nu,BB} \sim \nu^{2}$ is much lower, it could be 
       still difficult to catch prompt optical flares. Current radio sensitivities of ~50
       $\mu$Jy are already adequate for detecting reverse and forward shock peaks 
       but with predicted sensitivities as low as 2.3 $\mu$Jy in a 2-hour integration 
       (Chandra et al. 2010) all radio light curves in our sample would be easily 
       observed from early to late time with instruments such as the EVLA and ALMA. 
       
\end{itemize}

\section*{Acknowledgments} 
We thank the anonymous referee for valuable comments and suggestions that improved 
the paper. AM acknowledges funding from the Particle Physics and Astronomy Research
Council (PPARC). CGM is grateful for financial support from the Royal Society 
and Research Councils (UK). AG acknowledges founding from the Slovenian 
Research Agency and from the Centre of Excellence for Space Sciences and 
Technologies SPACE-SI, an operation partly financed by the European Union, 
European Regional Development Fund and Republic of Slovenia, Ministry of 
Higher Education, Science and Technology. IdG is partially supported by
Ministerio de Ciencia e Innovaci\'on (Spain), grant AYA2008-06189-C03
(including FEDER funds), and by Consejer\'{i}a de Innovaci\'on, Ciencia
y Empresa of Junta de Andaluc\'{i}a (Spain).The Liverpool Telescope is
operated by Liverpool John Moores University at the Observatorio del
Roque de los Muchachos of the Instituto de Astrofisica de Canarias. The
Faulkes Telescopes, now owned by Las Cumbres Observatory, are operated
with support from the Dill Faulkes Educational Trust. This work is
partially based on observations carried out with the IRAM Plateau de Bure 
Interferometer and observations collected at the German-Spanish Astronomical 
Center, Calar Alto, jointly operated by the Max-Planck-Institut f\"ur Astronomie 
Heidelberg and the Instituto de Astrof\'{i}sica de Andaluc\'{i}a (CSIC). IRAM is 
supported by INSU/CNRS (France), MPG (Germany) and IGN (Spain). We 
thank Calar Alto Observatory for allocation of director's discretionary time to this 
program. We also would like to thank M.R Zapatero-Osorio for the acquisition and
reduction of NOT data.The research of JG and AJCT is suppported by the
Spanish programmes AYA2007-63677, AYA2008-03467/ESP and
AYA2009-14000-C03-01. This work made use of data supplied by the UK {\it
Swift} Science Data Centre at the University of Leicester.

\clearpage

\clearpage
 
\clearpage
 
 \clearpage
 
\clearpage 
 
 \clearpage
 
 \clearpage

 \clearpage

\clearpage
\setcounter{table}{0}
\begin{landscape}
\begin{table*} \begin{center}
 \caption{Observed magnitudes and fluxes for the optical/infrared afterglow of GRB 090313. $\Delta t$ is the delay since the burst event in the observer frame. Magnitudes are not corrected  for Galactic absorption while $F_{\nu}$ are the absorption corrected converted flux densities. } \label{obslog0} \scriptsize
 \begin{tabular}{@{}cccccc|ccccccc}
\hline 
\hline
$\Delta t$ &  t$_{exp}$ & Filt. & Mag & $F_{\nu}$ & Ref. &$\Delta t$ &  t$_{exp}$ & Filt. & Mag & $F_{\nu}$ & Ref. \\
 (min) & (s) & & & (mJy) & & (min) & (s) & & & (mJy) &  \\
\hline \hline
 3.41 & 30.0 & $R_{\rm C}$ & $18.04 \pm 0.28$ & $0.228 \pm 0.062$ & FTN  & 7.30 & 10.0 & $i'$ & $16.11 \pm 0.10$ & $1.398 \pm 0.129$ & FTN \\
 9.89 &  30.0 &  $R_{\rm C}$ & $16.28 \pm 0.16$ & $1.196 \pm  0.177$ & FTN & 11.33 & 30.0 & $i'$ & $15.66 \pm 0.08$ & $2.117 \pm 0.129$ & FTN \\
14.71 &  60.0  & $R_{\rm C}$ & $16.07 \pm 0.10$ & $1.751 \pm 0.134$ & FTN & 16.71 & 60.0 & $i'$ & $15.31 \pm 0.08$ & $2.922 \pm 0.129$ & FTN \\
21.88 &  120.0  & $R_{\rm C}$ & $15.67 \pm 0.10$ & $1.897 \pm 0.193$ & FTN & 24.95 & 120.0 & $i'$ & $15.40 \pm 0.05$ & $2.689 \pm 0.129$ & FTN \\
32.21 &  180.0  & $R_{\rm C}$ & $16.02 \pm 0.11$ & $1.419 \pm 0.154$ & FTN & 26.20 & 180.0 & $i'$ & $15.68 \pm 0.05$ & $2.078 \pm 0.129$ & FTN \\
89.28 &  30.0  & $R_{\rm C}$ & $17.62 \pm 0.20$ & $0.348  \pm   0.065$ & FTN& 93.04 & 10.0 & $i'$ & $17.03 \pm 0.17$ & $0.599 \pm 0.129$ & FTN \\
95.70 &  30.0  & $R_{\rm C}$ & $17.47 \pm 0.17$ & $0.399  \pm 0.063$ & FTN & 97.10 & 30.0 & $i'$ & $16.79 \pm 0.10$ & $0.747 \pm 0.129$ & FTN \\
100.54 &  60.0  & $R_{\rm C}$ & $16.92 \pm 0.13$ &  $0.663  \pm 0.079$ & FTN & 102.44 & 60.0 & $i'$ & $16.65 \pm 0.08$ & $0.850 \pm 0.129$ & FTN \\
107.70 &  120.0  & $R_{\rm C}$ & $17.51 \pm 0.11$ &  $0.385 \pm  0.039$ & FTN & 110.56 & 120.0 & $i'$ & $17.09 \pm 0.09$ & $0.567 \pm 0.129$ & FTN \\
117.78 &  180.0  & $R_{\rm C}$ & $17.78 \pm 0.14$ &  $0.300 \pm   0.039$ & FTN & 121.75 & 180.0 & $i'$ & $17.43 \pm 0.13$ & $0.415 \pm 0.129$ & FTN \\
127.19 &  120.0  & $R_{\rm C}$ & $18.20 \pm 0.22$ &   $0.204 \pm  0.042$ & FTN & 130.82 & 120.0 & $i'$ & $17.10 \pm 0.11$ & $0.562 \pm 0.129$ & FTN \\
137.99 &  180.0  & $R_{\rm C}$ & $17.99 \pm 0.18$ &  $0.247  \pm  0.041$ & FTN & 141.99 & 180.0 & $i'$ & $17.18 \pm 0.10$ & $0.522 \pm 0.129$ & FTN \\
204.80 &  300.0  & $R_{\rm C}$ & $18.14 \pm 0.10$ &  $0.216  \pm  0.020$ & FTN & 169.90 & 900.0 & $i'$ & $17.21 \pm 0.10$ & $0.508 \pm 0.129$ & FTN \\
210.18 &  300.0  & $R_{\rm C}$ & $18.23 \pm 0.10$ & $0.198  \pm  0.018$ & FTN & 186.03 & 900.0 & $i'$ & $17.78 \pm 0.13$ & $0.300 \pm 0.129$ & FTN \\
215.56 &  300.0 & $R_{\rm C}$ & $18.49 \pm 0.12$ & $0.156 \pm 0.017$ & FTN & 336.59 & 300.0 & $i'$ & $17.21 \pm 0.15$ & $0.508 \pm 0.129$ & FTN \\
220.93 &  300.0 & $R_{\rm C}$ & $18.43 \pm 0.10$ & $0.165  \pm   0.015$ & FTN & 352.72 & 1500.0 & $i'$ & $17.26 \pm 0.16$ & $0.485 \pm 0.129$ & FTN \\
226.31 &  300.0  & $R_{\rm C}$ & $18.37 \pm 0.12$ & $0.174  \pm   0.019$ & FTN & 86288.09 & 600.0 & $i'$ & $20.94 \pm 0.22$ & $0.0163 \pm 0.0034$ & FTN \\
231.69 &   300.0 & $R_{\rm C}$ & $17.53 \pm 0.15$ & $0.378 \pm    0.052$ & FTN & 132306.85 & 1800.0 & $i'$ & $21.12 \pm 0.10$ & $0.0136 \pm 0.0012$ & FTN \\
306.69 &   900.0 & $R_{\rm C}$ & $17.40 \pm 0.10$ & $0.426  \pm   0.039$ & FTN & 865.36 & 900.0 & $i'$ & $19.01 \pm 0.11$ & $0.097 \pm 0.010$ & LT \\
322.82 &  900.0  & $R_{\rm C}$ & $17.27 \pm 0.15$ & $0.480  \pm   0.066$ & FTN & 880.96 & 900.0 & $i'$ & $18.73 \pm 0.08$ & $0.125 \pm 0.009$ & LT \\
1407.83 & 1800.0  & $R_{\rm C}$ & $19.85 \pm 0.11$ & $0.0446  \pm  0.005$ & FTN & 1010.82 & 1800.0 & $i'$ & $19.29 \pm 0.25$ & $0.074 \pm 0.017$ & LT \\
86255.28 &  600.0  & $R_{\rm C}$ & $21.58 \pm 0.15$ & $0.0091 \pm  0.0013$ & FTN & 1156.30 & 900.0 & $i'$ & $18.86 \pm 0.05$ & $0.111 \pm 0.005$ & LT \\
86265.98 &  600.0  & $R_{\rm C}$ & $21.63 \pm 0.13$ & $0.0086 \pm   0.0010$ & FTN & 1169.30 & 900.0 & $i'$ & $18.84 \pm 0.07$ & $0.113 \pm 0.007$ & LT \\
86277.15 &  600.0  & $R_{\rm C}$ & $21.75 \pm 0.15$ & $0.0077 \pm   0.0010$ & FTN & 2431.86 & 7200.0 & $i'$ & $20.25 \pm 0.07$ & $0.0309 \pm 0.0020$ & LT \\
131244.99 & 1200.0 & $R_{\rm C}$ & $21.40 \pm 0.64$ &  $0.0107 \pm   0.0067$ & FTN & 3870.60 & 7200.0 & $i'$ & $20.43 \pm 0.08$ & $0.0261\pm 0.0020$ & LT \\
828.90 & 1800.0 & $r'$ &  $19.73 \pm 0.10$ &  $0.0498  \pm  0.0046$ & LT & 27001.76 & 600.0 & $i'$ & $21.09 \pm 0.15$ & $0.0142 \pm 0.0020$ & LT\\
983.50 &  3600.0  & $r'$ &  $18.65 \pm 0.11$ &  $  0.1348   \pm 0.0137$ & LT & 4872.80 & 1800.0 & $i'$ & $21.23 \pm 0.18$ & $0.0125 \pm 0.0020$ & FTS \\
1200.25 & 1800.0  & $r'$ &  $19.60 \pm 0.23$ &  $ 0.0562   \pm 0.0120$ & LT & 6285.20 & 1800.0 & $i'$ & $21.10 \pm 0.20$ & $0.0138 \pm 0.025$ & FTS \\
2507.32 & 7200.0   & $r'$ &  $20.76 \pm 0.11$ &  $ 0.0193  \pm 0.0019$ & LT & 811.10 & 900.0 & I & $18.75 \pm 0.17$ & $0.123 \pm 0.020$ & OSN \\
3933.26 &  7200.0  & $r'$ &  $21.47 \pm 0.10$ &  $0.0100  \pm  0.0010$ & LT& 829.13 & 900.0 & I & $18.62 \pm 0.10$ & $0.139 \pm 0.013$ & OSN  \\
26990.86 & 600.0 & $r'$ & $21.62 \pm 0.21$ & $0.0087 \pm 0.0017$ & LT & 844.51 & 900.0 & I & $18.64 \pm 0.10$ & $0.136 \pm 0.012$ & OSN \\
4839.42 & 1800.0 & $R_{\rm C}$ & $21.64 \pm 0.20$ & $0.0086 \pm 0.0016$ & FTS & 855.73 & 900.0 & I & $18.88 \pm 0.12$ & $0.109 \pm 0.012$ & OSN \\
6251.83 & 1800.0  & $R_{\rm C}$ & $21.44 \pm 0.15$ & $0.0103 \pm 0.0014$ & FTS & 868.40 & 800.0 & I & $18.32 \pm 0.15$ & $0.182 \pm 0.025$ & OSN \\
935.04 & 600.0 & $R_{\rm C}$ & $19.31 \pm 0.12$ & $0.0734 \pm 0.0081$ & IAC80 & 1203.40 & 400.0 & I & $19.23 \pm 0.15$ & $0.079 \pm 0.011$ & OSN \\
1167.45 & 1380.0 & $R_{\rm C}$ & $19.74 \pm 0.26$ & $0.0494 \pm 0.0119$ & IAC80 & 2239.18 & 1000.0 & I & $20.15 \pm 0.15$ & $0.033 \pm 0.005$ & OSN \\
9695.8 & 13800.0 &  $R_{\rm J}$ & $> 23.7$ & --- & 1.23Caha & 1885.18 & 10260.0 & I & $19.77 \pm 0.26$ & $0.048 \pm 0.011$ & Mitsume \\
11171.1 & 9600.0 & $R_{\rm J}$ & $> 23.5$ & --- & 1.23Caha & 1173.80 & 540.0 & J & $17.45 \pm 0.20$ & $0.1708 \pm 0.0316$ & NOT \\
1152.38 & 1080.0 & K & $15.33 \pm 0.25$ & $0.495 \pm 0.115$ & NOT & 8300.50 & 3600.0 & J & $19.27 \pm 0.12$ & $0.0319 \pm 0.0035$ & 3.5Caha \\
\hline
 \end{tabular}
\end{center}
\end{table*}
\end{landscape}

\clearpage
\setcounter{table}{1}
\begin{table*} \begin{center}
 \caption{Detected fluxes for the afterglow of GRB 090313 in the radio and mm bands.} \label{obslog1} \scriptsize
 \begin{tabular}{@{}cccc}
\hline 
\hline
$\Delta t$ &  Frequency Range & $F_{\nu}$ & Ref. \\ 
 (days) & (GHz) & (mJy) &  \\ 
\hline 
\hline
1.01 & 92.5 & $4.00 \pm 0.60$ & CARMA, GCN 9005 \\ 
1.71 & 4.9 & $0.026 \pm 0.038$ & WSRT, GCN 9000 \\
2.75 & 14.5-17.5 & $0.800 \pm 0.080$ & AMI \\
3.68 & 14.5-17.5 & $0.882 \pm 0.077$ & AMI \\
4.57 & 105.38 & $1.681 \pm 0.153$ & PdBI \\
4.67 & 14.5-17.5 & $0.815 \pm 0.129$ & AMI \\
5.63 & 228.00 & $0.605 \pm 0.507$ & PdBI \\
5.85 & 8.46 & $0.269 \pm 0.031$ & VLA, GCN 9011 \\
6.78 & 14.5-17.5 & $0.718 \pm 0.097$ & AMI \\
7.40 & 4.9 & $0.165 \pm 0.030$ & WSRT, GCN 9016 \\
9.67 & 14.5-17.5 & $0.655 \pm 0.069$ & AMI \\
13.59 & 87.205 & $0.666 \pm 0.126$ & PdBI \\
16.74 & 14.5-17.5 & $0.569 \pm 0.140$ & AMI \\
19.54 & 110.00 & $-0.206 \pm 0.304$ & PdBI \\
36.66 & 14.5-17.5 & $0.427 \pm 0.100$ & AMI \\
46.63 & 14.5-17.5 & $0.080 \pm 0.130$ & AMI \\
\hline
 \end{tabular}
\end{center}
\end{table*}

\clearpage
 \setcounter{table}{2}
\begin{table*} \begin{center}
 \caption{Fit results of the spectral energy distributions of GRB 090313. t$_{\rm rf}$ is the time of the SED in the rest frame while $t_{\rm obs}$ is the corresponding time in the observed frame.} \label{tabsed} 
 \begin{tabular}{@{}ccccc}
\hline 
\hline
SED &  t$_{\rm rf}$ & $t_{\rm obs}$ & $\beta_{\rm O}$ & \\
  & (s) &  (s) &\\
\hline
1 & 100 & 437.5 & $1.19 \pm 0.02$ \\ 
2 & 600 & 2625 & $1.21 \pm 0.05$ \\ 
3 & 2000 & 8750& $1.39 \pm 0.07$ \\
4 & $1.6 \times 10^{4}$ & $7 \times 10^{4}$ & $1.35 \pm 0.30$ \\
\hline
 \end{tabular}
\end{center}
\end{table*}

\clearpage
  \setcounter{table}{3}
  \begin{landscape}
\begin{table*} 
\centering
 \caption{GRBs with detected optical peaks. All the redshifts are spectroscopically confirmed but the redshift for GRB 070420, for which the pseudo-z from [15] and the photo-z from [6] are reported. References:[1] Akerlof et al. 1999; [2] Galama et al. 1999; [3] Bersier  et al. 2006; [4] Sakamoto et al. 2004; [5] Rykoff et al. 2004; [6] Oates et al. 2009; [7] Perri et al. 2007; [8] Cenko et al. 2006; [9] Vestrand et al. 2006; [10] Molinari et al. 2007; [11] Rykoff et al. 2009; [12] Mundell et al. 2007a; [13] Schady et al. 2007; [14] Melandri et al. 2009; [15] Klotz et al. 2008; [16] Kruhler et al. 2009a; [17] Greiner et al. 2009; [18] Guidorzi et al. 2009; [19] Guidorzi et al. 2009b; [20] Falcone et al. 2006; [21] Page et al. 2006; [22] Stratta et al. 2009; [23] Page et al. 2009; [24] Golenetskii et al. 2007; [25] Kruhler et al. 2009b. Where possible, the value of E$_{\rm iso}$ has been taken from Amati et al. (2008) and Rossi et al. (2008)} \label{tabprop} \tiny
 \begin{tabular}{@{}ccccccccccccccc}
\hline 
\hline
GRB &  $\alpha_{\rm rise}$ & $\alpha_{\rm decay}$ & t$_{\rm peak}$ & $\alpha_{\rm X}$ & A$_{\rm V}$ & T$_{90}$ & z & $\Gamma$ & $\epsilon_e$ & F$_p$ & E$_{\rm iso}$ & Ref.\\
  & & & (s) & & mag & (s) & & & Upper Limit & (mJy) & (10$^{52}$ erg) &\\
\hline
990123 & $> -3.5$ & $2.1 \pm 0.2$ & $50 \pm 15$ & $1.46 \pm 0.04$ & 0.053 & $63.3 \pm 0.3$ & 1.61 & $\approx 450$ & 0.004 & & $229 \pm 37$ & [1],[2]\\
020903 & $-4.6 \pm 3.7$ & $1.8 \pm 3.0$ & $\sim 10^{5}$ & & 0.098 & $9.8
			  \pm 0.6$ & 0.25 & & 18.466 & $0.10 \pm 0.05$ &
					      $ (2.4 \pm 0.6)\times10^{-3}$ & [3],[4] \\
030418 & $-0.62 \pm 0.13$ & $1.34 \pm 0.06$ & $2401 \pm 303$ & & 0.077 & $135 \pm 5$ & $<5$ & & & $0.40 \pm 0.10$ & & [5]\\ 
050730 & $0.15 \pm 0.50$ & $0.89 \pm 0.05$ & $\sim 750$ & $0.44^{+0.14}_{-0.08}$ & 0.155 & $155 \pm 20$ & 3.97 & $\approx 110$ & 0.043 & $0.57 \pm 0.10$ &$\sim 8 $ & [6],[7]\\
050820A & $-0.35 \pm 0.02$ & $0.97 \pm 0.01$ & $\sim 600$ & $0.93 \pm 0.03$ & 0.137 & $26 \pm 2$ & 2.615 & $\approx 145$ & 0.021 & $4.05 \pm 0.15$ & $97.4 \pm 7.8$ & [8],[9]\\
060418 & $-2.7^{+1.0}_{-1.7}$ & $1.28 \pm 0.05$ & $153 \pm 10$ & $1.4 \pm 0.1$ & 0.702 & $52 \pm 1$ & 1.489 & $\approx 165$ & 0.014 & $8.00 \pm 0.50 $ & $13 \pm 3$ & [10],[20]\\
060605 & $-0.84 \pm 0.12$ & $1.16 \pm 0.06$ & $475 \pm 53$ & $0.34 \pm 0.08$ & 0.159 & $79.1 \pm 3.0$ & 3.78 & $\sim 110$ & 0.042 & $2.00 \pm 0.10$ & $2.5^{+3.1}_{-0.6}$ & [11]\\
060607A & $-3.6^{+0.8}_{-1.1}$ & $1.27^{+0.16}_{-0.11}$ & $180^{+5}_{-6}$ & $1.09 \pm 0.04$ & 0.089 & $100 \pm 5$ & 3.082 & $\approx 180$ & 0.014 & & $11$ & [10],[21]\\
060904B & $-1.7^{+2.0}_{-0.7}$ & $1.11^{+0.14}_{-0.20}$ & $479.4^{+24}_{-15}$ & $0.76 \pm 0.04$ & 0.549 & $172 \pm 5$ & 0.703 & $\sim 60$ & 0.093 & $0.58 \pm 0.08$ & $0.30^{+0.19}_{-0.06}$ & [11]\\
   & $-0.82 \pm 0.15$ & $1.00 \pm 0.18$ & $\sim 550.0$ & $0.86 \pm 0.05$ & & & & & & & & [15]\\
061007 & $\sim -9.0$ & $1.7 \pm 0.1$ & $70 \pm 15$ & $1.68 \pm 0.02$ & 0.024 & $75.3 \pm 5.0$ & 1.261 & $\sim 265$ & 0.005 & $280.0 \pm 10.0$ & $86 \pm 9$ & [12],[13],[11]\\
070419A & $-1.56 \pm 0.70$ & $0.61 \pm 0.09$ & $450 \pm 20$ & $1.27^{+0.18}_{-0.12}$ & 0.087 & $115.6 \pm 5.0$ & 0.97 & $\sim 60$ & 0.100 & $0.15 \pm 0.05$ & $0.16$ & [14]\\
070420 & $-1.26 \pm 0.19$ & $0.88 \pm 0.09$ & $\sim 200 \pm 10$ & $0.23 \pm 0.05$ & 1.561 & $76.5 \pm 4.0$ & $1.56 \pm 0.35$ & & & & & [15], [24]\\
  & $-0.73 \pm 0.14$ & $1.67 \pm 0.15$ & $\sim 350.0$ & & & & $3.01^{+0.96}_{-0.68}$ & $\approx 230$ & & & & [6]\\
071031 & $-0.7 \pm 0.1$ & $0.97 \pm 0.06$ & $\sim 1000$ & $0.99 \pm 0.12$ & 0.036 & $180 \pm 10$ & 2.692 & $\sim 70$ & 0.102 & $0.50 \pm 0.05$ & $\sim 1.5$ & [16] \\
080129 & $-1.35 \pm 0.15$ & $0.5 \pm 0.1$ & $\sim 12000$ & $1.5 \pm 0.1$ & 3.046 & $48 \pm 2$ & 4.394 & $\approx 45$ & 0.355 & $0.25 \pm 0.10$ & $\sim 7$ & [17],[22]\\
080330 & $-0.4 \pm 0.2$ & $2.02 ^{+0.85}_{-0.75}$ & $\sim 600$ & $0.26 \pm 0.10$ & 0.051 & $61 \pm 9$ & 1.51 & $< 60$ & 0.109 & $0.35 \pm 0.05$ & $0.21 \pm 0.05$ & [18]\\
080603A & $-3.2 \pm 0.3$ & $1.2 \pm 0.1$ & $\sim 1600$ & $1.0 \pm 0.1$ & 0.138 & $180 \pm 10$ & 1.688 & $\sim 55$ & 0.137 & $0.18 \pm 0.05$ & $1.5 \pm 0.5$ & [19]\\
080710 & $-1.11 \pm 0.03$ & $0.63 \pm 0.02$ & $1829 \pm 19$ & $1.0 \pm 0.2$ & 0.230 & $120 \pm 17$ & 0.845 & $\sim 40$ & 0.213 & $0.90 \pm 0.10$ & $\sim 0.56$ & [25]\\
080810 & $-1.32 \pm 0.11$ & $1.22 \pm 0.09$ & $\sim 100$ & $1.81 \pm 0.20$ & 0.087 & $108 \pm 5$ & 3.355 & $\sim 260$ & 0.007 & $0.003 \pm 0.001$ & $\sim 30$ & [23]\\
090313 & $-1.72 \pm 0.41$ & $1.25 \pm 0.08$ & $1061 \pm 154$ & $0.83 \pm 0.49$ & 0.090 & $78 \pm 19$ & 3.375 & $\sim 80$ & 0.074 & $1.80 \pm 0.20$ & $\sim 3.2$ & This work\\
\hline
 \end{tabular}
\end{table*}
\end{landscape}

\end{document}